\documentclass[notitlepage,aps,prl,showpacs,onecolumn,superscriptaddress,nofootinbib]{revtex4-1}

\pdfoutput=1 

\usepackage{mathtools}
\usepackage{amssymb}
\usepackage{xcolor}
\usepackage{graphicx}
\usepackage[caption=false]{subfig}
\usepackage{verbatim}
\usepackage{bm}
\usepackage{bbold}
\usepackage{xr}

\usepackage{bibunits} 
\def\bibsection{\section*{References}}  

\usepackage[symbol]{footmisc}

\newcommand\numberthis{\addtocounter{equation}{1}\tag{\theequation}}
\newcommand\mb{\mathbf}
\newcommand{\mbs}{\boldsymbol}
\newcommand{\rot}{\mbs{\mathcal{R}}}
\newcommand{\St}{\mathcal{S}}
\newcommand{\Id}{\mathbb{1}}

\newcommand{\G}{\mathcal{G}}

\begin{document}

\begin{bibunit}[rsc]

\title{Van't Hoff's law for active suspensions: 
the role of the solvent chemical potential}
\author{Jeroen Rodenburg}
\email{A.J.Rodenburg@uu.nl}
\affiliation{Institute for Theoretical Physics, Center fer Extreme Matter and Emergent Phenomena, Utrecht University, Princetonplein 5, 3584 CC Utrecht, The Netherlands}
\author{Marjolein Dijkstra}
\affiliation{Soft Condensed Matter Group, Debye Institute for Nanomaterials Science, Utrecht University, Princetonplein 5, 3584 CC Utrecht, The Netherlands}
\author{Ren\'e van Roij}
\affiliation{Institute for Theoretical Physics, Center fer Extreme Matter and Emergent Phenomena, Utrecht University, Princetonplein 5, 3584 CC Utrecht, The Netherlands}

\begin{abstract}
We extend Van't Hoff's law for the osmotic pressure to a suspension of active Brownian particles. The propelled particles exert a net reaction force on the solvent, and thereby either drive a measurable solvent flow from the connecting solvent reservoir through the semipermeable membrane, or increase the osmotic pressure and cause the suspension to rise to heights as large as micrometers for experimentally realized microswimmers described in the literature. The increase in osmotic pressure is caused by the background solvent being, in contrast to passive suspensions, no longer at the chemical potential of the solvent reservoir. The difference in solvent chemical potentials depends on the colloid-membrane interaction potential, which implies that the osmotic pressure is a state function of a state that \emph{itself} is influenced by the membrane potential.
\end{abstract}

\maketitle

\twocolumngrid

\section*{Introduction}
\begin{figure}
 \includegraphics[width=\linewidth]{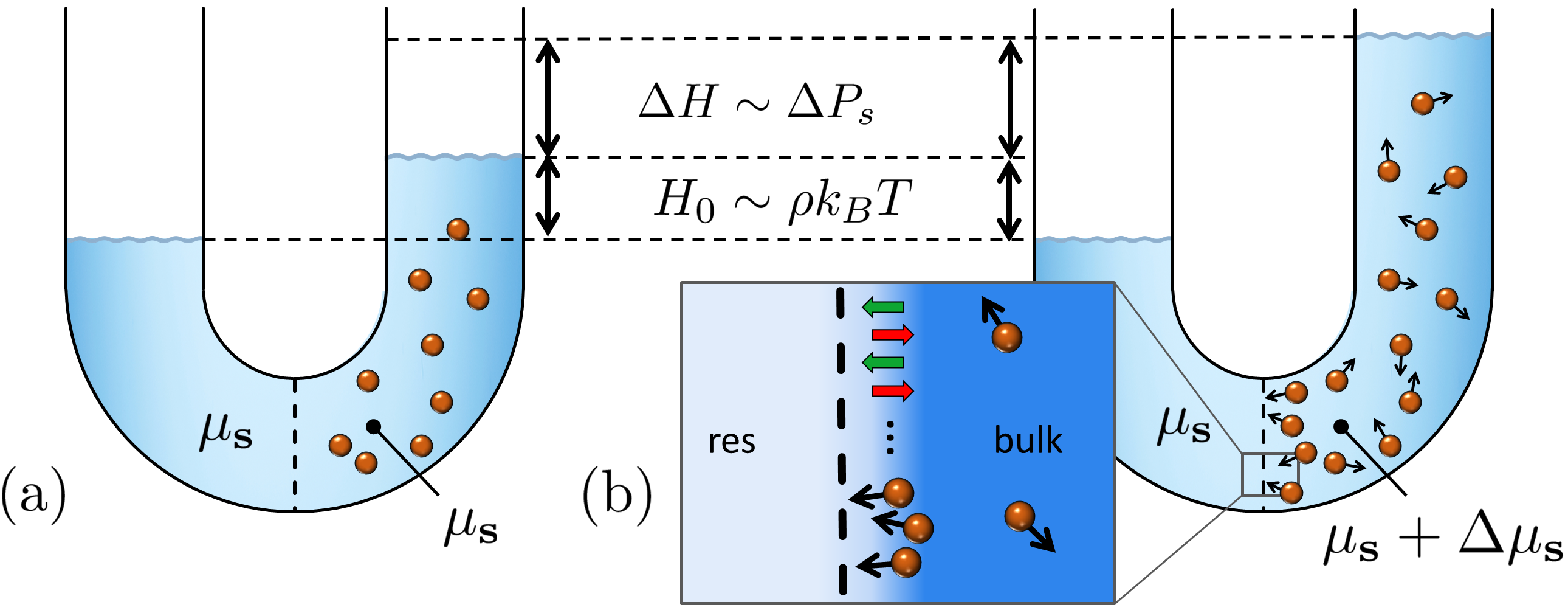}
 \caption{Schematic setup to measure osmotic pressure $\Pi$ from the height difference $H = H_0 + \Delta H$ between the two menisci. (a) For a passive system, the solvent chemical potential of the suspension equals the reservoir chemical potential $\mu_s$, such that $\Pi = \rho k_BT$ and $H = H_0 \sim \rho k_BT$. (b) For an active system, colloids tend to `propel into' the membrane (green arrows), thereby exerting the opposite reaction force on the solvent (red arrows). As a result, the solvent pressure and solvent chemical potential in the bulk suspension increase by $\Delta P_s$ and $\Delta \mu_s$, respectively, indicated by the darker blue background, such that $\Pi$ and $H$ increase by $\Delta P_s$ and $\Delta H \sim \Delta P_s$, respectively.}
 \label{fig:NicePicture}
\end{figure}

\begin{figure}[t]
\center
 \includegraphics[width=.9\linewidth]{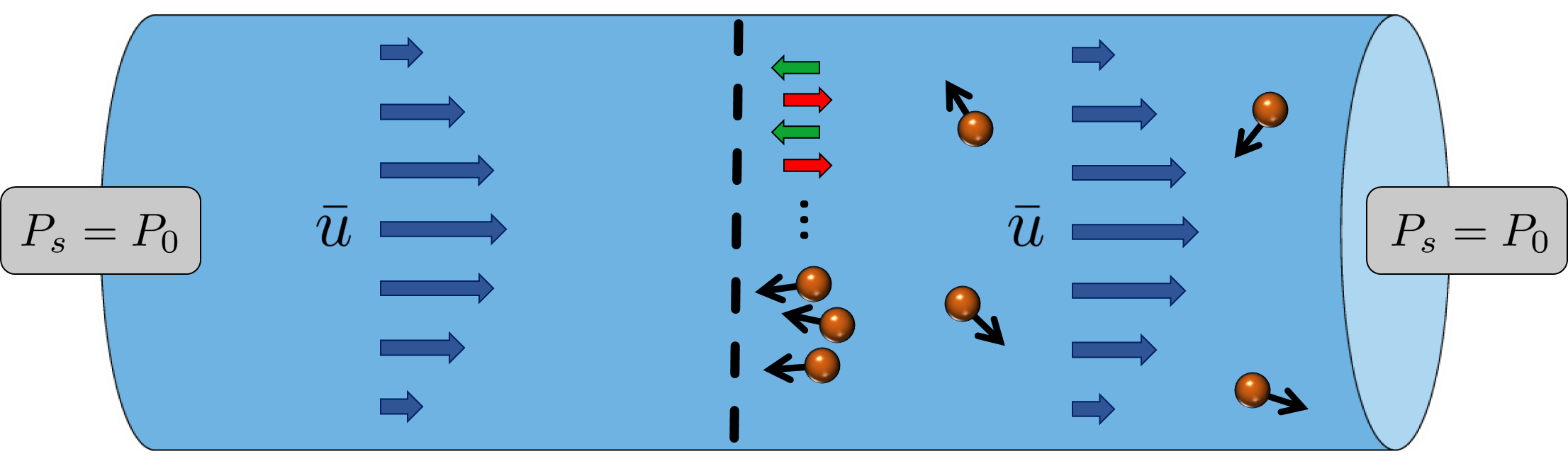}
 \caption{Schematic setup to measure how active colloids confined to one side of a semipermeable membrane in an open pipe affect the solvent, viewed in the lab frame. As the colloids tend to `propel into' the membrane (green arrows), they exert the opposite reaction force on the solvent (red arrows). Under boundary conditions of equal solvent pressure $P_s$ on either side of the pipe, the reaction force drives a parabolic solvent flow profile of mean velocity $\bar{u}$, as indicated by the blue arrows.}
 \label{fig:PipeFlow}
\end{figure}

\noindent In 1887, Van't Hoff formulated his famous law stating that the osmotic pressure $\Pi$ of a dilute suspension equals the pressure $\rho k_BT$ of a dilute gas of the same concentration $\rho$ and temperature $T$, with $k_B$ the Boltzmann constant \cite*{Hoff1,Hoff2,Hoff3}. In Van't Hoff's interpretation, the total pressure of the suspension $P_{\text{tot}}(\rho,\mu_s) = \rho k_B T + P_s(\mu_s)$ decomposes into the sum of the effective colloid-only pressure $\rho k_BT$ and a `background' pressure $P_s(\mu_s)$ of the solvent at chemical potential $\mu_s$. 
In the typical experimental setup to measure osmotic pressure (Fig. \ref{fig:NicePicture}), $\mu_s$ is set by a solvent reservoir that connects to the suspension via a membrane permeable to solvent only. The net force per unit area exerted on the membrane defines the osmotic pressure, and results from the difference in suspension pressure $P_{\text{tot}}(\rho,\mu_s)$ and reservoir pressure $P_s(\mu_s)$. As this pressure difference induces a height difference $H$ between the two menisci, the osmotic pressure $\Pi \sim H$ can be directly inferred.\\
	\indent Van't Hoff's law does not apply to non-equilibrium suspensions of active particles that constantly convert energy into directed motion, such as swimming bacteria \cite{EColiBook,PoonEColi} or artificial microswimmers \cite{ReviewMicroSwimmers}. Not only are these systems promising for applications in e.g. self-assembly \cite{SelfAssembly1, SelfAssemblyVassilis} and targeted cargo transport \cite{Cargo1,Cargo2}, they also display remarkable phase behaviour \cite{MIPSReview,MarchettiABPReview,Theurkauff2012,PalacciLivingCrystals,ExpMIPS,VicsekModel,MarchettiMIPS,Redner2013,Bialke2013,Wysocki2014} that calls for an underpinning thermodynamic framework \cite{Lowen2014,BradyTowardsThermodynamics,Brader2015,BradyMixesActiveParticles,MarconiMaggi,ActiveH,BradyReview,ActiveEntropyProduction,Siddharth2016,TailleurGeneralThermo,EffectiveEquilibrium2017,Berend2016}. An essential prerequisite for such a framework is that thermodynamic properties can be expressed as functions of variables that characterize the system state, a seemingly trivial condition that was nonetheless questioned for the osmotic pressure \cite{Cates}.\\
\indent Previous studies of the pressure \cite{MarchettiSwimPressure,Wittkowski2014,BradySwimPressure,BradyAppliesExternalField,EoSExperiment,CatesInteractingSpheres,MarconiMaggi2,Cates,BradyBodyForce,Winkler,Lowen,BradyForceOnBoundary,UnderdampedDumbells,SpeckJack,BradySupernova,Nikola2016,Falasco,StarkPhaseSeparation,Falasco2016,Levis2017,Junot2017,BradyInertia} mostly addressed self-propelled particles on a substrate, or equivalently, an effective colloid-only system without an explicit solvent. The solvent \emph{was} explicitly modelled in Ref. \cite{Lion2014}, but only as a passive species unaffected by the propulsion force. However, as the propulsion force is internal \cite{MarchettiReview}, the solvent - and in particular its pressure - \emph{is} affected by the opposite reaction force \cite{BradyBodyForce}.\\
	\indent In this work, we apply this insight to extend Van't Hoff's law to active suspensions. We show the osmotic pressure to increase with activity due to a difference in \emph{solvent} pressure that develops between the suspension and the reservoir. The effect of this solvent pressure difference is predicted to be experimentally measurable, either as an additional meniscus rise $\Delta H$ (Fig. \ref{fig:NicePicture}), or as a solvent flow through a semipermeable membrane towards the active particles in an open system (Fig. \ref{fig:PipeFlow}). The solvent pressure difference implies also a difference in solvent chemical potential, that, remarkably, depends on the details of the colloid-membrane interactions. We will conclude that the osmotic pressure \emph{is} a state function of a state that \emph{itself}, however, is affected by the colloid-membrane interaction potential.\\
\section*{Model}
\noindent We model the effective one-component system of suspended particles as overdamped active Brownian particles (ABPs) \cite{ABPs,MarchettiABPReview}. Every particle is characterized by a three-dimensional position $\mb{r}$ and an orientation $\mb{\hat{e}}$. It is well known that the probability density $\psi(\mb{r},\mb{\hat{e}},t)$ satisfies the Smoluchowski equation (see section 1 of the ESI\footnote[2]{The Electronic Supplementary Information (ESI) is included in this document after the main paper.})
\begin{equation*}
 \label{eqn:FP}
  \partial_t \psi(\mb{r},\mb{\hat{e}},t)
   =
  - \mbs{\nabla} \cdot \mb{j}(\mb{r},\mb{\hat{e}},t)
  - \mbs{\nabla}_{\mb{\hat{e}}} \cdot \mb{j}_{\mb{\hat{e}}}(\mb{r},\mb{\hat{e}},t),
 \numberthis
\end{equation*}
\noindent and that the translational flux $\mb{j}$ and rotational flux $\mb{j}_{\mb{\hat{e}}}$ follow from the force and torque balance,
\begin{equation*}
 \label{eqn:FPCurrents}
  \begin{alignedat}{1}
   0 &=
   -\gamma_t \mb{j}_{\phantom{\mb{\hat{e}}}}
   -\psi\mbs{\nabla} V(\mb{r},\mb{\hat{e}})
   + \gamma_t v_0 \psi \mb{\hat{e}}
   - k_BT \mbs{\nabla}\psi
     \quad \text{and}\\
   0 &=
   -\gamma_r\mb{j}_{\mb{\hat{e}}} 
   -\psi\mbs{\nabla}_{\mb{\hat{e}}} V(\mb{r},\mb{\hat{e}}) 
   - k_BT \mbs{\nabla}_{\mb{\hat{e}}}\psi
   ,
  \end{alignedat}
 \numberthis
\end{equation*}
\noindent respectively, between (i) the frictional force and torque, with friction coefficients $\gamma_t$ and $\gamma_r$, (ii) an external force and torque generated by the external potential $V(\mb{r},\mb{\hat{e}})$ acting on every particle, (iii) a constant self-propulsion force, corresponding to propulsion speed $v_0$, acting along each particle's orientation $\mb{\hat{e}}$, and (iv) Brownian forces and torques giving rise to translational and rotational diffusion. 
In order to focus on the essential physics, we follow Van't Hoff and consider the dilute limit, where effective colloid-colloid interactions can be ignored, and where also hydrodynamic interactions are expected to be nonessential. Furthermore, we assume a steady state, i.e. $\partial_t\psi=0$. We analyse the force balance by taking the zeroth moment of Eq. (\ref{eqn:FPCurrents}), which upon defining the density $\rho(\mb{r}) \equiv \int\mathrm{d}\mb{\hat{e}}\psi(\mb{r},\mb{\hat{e}})$, the polarization $\mb{m}(\mb{r}) \equiv \int\mathrm{d}\mb{\hat{e}}\psi(\mb{r},\mb{\hat{e}})\mb{\hat{e}}$, and the colloid flux $\mb{J}(\mb{r})\equiv\int\mathrm{d}\mb{\hat{e}}\mb{j}(\mb{r},\mb{\hat{e}})$, yields the balance
\begin{flalign*}
 \label{eqn:FBC}
   \mb{0} 
   &=
   - \gamma_t \mb{J}(\mb{r})    
   \! - \! \int \! \! \mathrm{d} \mb{\hat{e}} 
           \psi \mbs{\nabla} V(\mb{r},\mb{\hat{e}})
      + \gamma_t v_0 \mb{m}(\mb{r})
      - k_BT\mbs{\nabla} \rho(\mb{r})\\
    &\equiv
    \mb{f}^f(\mb{r}) + \mb{f}^e(\mb{r}) + \mb{f}^p(\mb{r}) - \mbs{\nabla} P(\mb{r})
 \numberthis
\end{flalign*}
\noindent between the frictional body force $\mb{f}^f$, the external body force $\mb{f}^e$, the propulsion body force $\mb{f}^p$, and the pressure gradient force $-\mbs{\nabla} P$. The form of the propulsion body force,
\begin{equation*}
 \label{eqn:fp}
  \mb{f}^p(\mb{r}) \equiv \gamma_tv_0 \mb{m}(\mb{r}),
 \numberthis
\end{equation*}
is easily understood as the sum of propulsion forces $\gamma_tv_0\mb{\hat{e}}$ acting on individual colloids. Just like the frictional force $\mb{f}^f$, the propulsion force $\mb{f}^p$ is an \emph{internal} force.\\
	\indent We now turn our attention to the solvent, that we assume to be incompressible and at small Reynolds number. On a scale coarse-grained over the colloids - i.e. the same scale Eq. (\ref{eqn:FBC}) applies to - the local solvent velocity $\mb{u}(\mb{r})$ is governed by the Stokes equation 
\begin{equation*}
 \label{eqn:FBS}
  \mb{f}^e_s(\mb{r}) - \mb{f}^f(\mb{r}) - \mb{f}^p(\mb{r}) - \mbs{\nabla}P_s(\mb{r}) + \eta \mbs{\nabla}^2 \mb{u}(\mb{r}) = 0,
 \numberthis
\end{equation*}
\noindent as derived in section 2 of the ESI$^{\dag}$.
Eq. (\ref{eqn:FBS}) is simply the solvent force balance equipped with a possible external body force $\mb{f}^e_s(\mb{r})$, and the opposite internal body forces $-\mb{f}^f(\mb{r})$ and $-\mb{f}^p(\mb{r})$ as reaction forces, in accordance with Newton's third law. Furthermore, $P_s(\mb{r})$ is the solvent pressure, and $\eta$ the dynamic viscosity.
\section*{Osmotic pressure}
\noindent To represent the setting of Fig. \ref{fig:NicePicture}, we assume an external potential due to a semipermeable membrane that is planar and normal to the Cartesian unit vector $\mb{\hat{z}}$, i.e. $V(\mb{r},\mb{\hat{e}}) = V(z,\theta)$, with $\cos\theta \equiv \mb{\hat{e}} \cdot \mb{\hat{z}}$. This implies $\psi(\mb{r},\mb{\hat{e}}) = \psi(z,\theta)$, $\mb{J}(\mb{r}) = J_z(z) \mb{\hat{z}}$ etc. The potential $V(z,\theta)$ is assumed to decay from $\infty$ in an infinitely large reservoir, located at $z<0$ and containing $z$-coordinate $z_{\text{res}} \ll 0$ in bulk, to $0$ in the suspension, located at $z > 0$ and containing $z_b \gg 0$ in bulk. The zeroth moment of Eq. (\ref{eqn:FP}), $\partial_z J_z(z) = 0$, together with a no-flux boundary condition, then implies $J_z(z) = 0$, and hence the frictional body force $\mb{f}^f(z) = 0$. For a state without any solvent flow ($\mb{u} = \mb{0}$), and for a membrane perfectly invisible to the solvent ($\mb{f}^e_s = \mb{0}$), Eq. (\ref{eqn:FBS}) then simplifies to $-f^p_z(z) - \partial_zP_s(z)=0$.\\
	\indent For a passive system, where the propulsion body force $f^p_z(z) = 0$, this solvent force balance guarantees equal solvent pressures in the bulk suspension and solvent reservoir, i.e. $\Delta P_s \equiv P_s(z_b) - P_s(z_{\text{res}}) =0$. In an active system, however, the existence of a nonzero propulsion force $f^p_z(z)$ results in a difference in these solvent pressures, derived in section 4 of the ESI$^{\dag}$ to be
\begin{equation*}
 \label{eqn:DeltaPs} 
  \begin{alignedat}{1}
 \Delta P_s 
  &= 
  - \int_{z_{\text{res}}}^{z_b} \! \! \! \! \mathrm{d}z f_z^p(z) \\
  &=
     \  \frac{\gamma_t\gamma_rv_0^2}{6k_BT} \rho
     \ - \frac{\gamma_tv_0}{2k_BT} \int_{z_{\text{res}}}^{z_b} \mathrm{d}z \int \mathrm{d}\mb{\hat{e}}
     \psi(z,\theta) \sin(\theta) \partial_{\theta}V(z,\theta).
  \end{alignedat}
 \numberthis
\end{equation*}
\noindent The first term on the right-hand side of Eq. (\ref{eqn:DeltaPs}) corresponds to what is known as the swim pressure \cite{MarchettiSwimPressure,BradySwimPressure, Winkler}, which we thus actually identify as a difference in solvent pressure. The second term on the right-hand side of Eq. (\ref{eqn:DeltaPs}), present for particles experiencing a torque $-\partial_{\theta}V(z,\theta)$, is of special interest because it leads to the conlusion that $\Delta P_s$ depends on the potential $V(z,\theta)$. This issue will be discussed later.\\
	\indent The force balance of the total suspension simply follows as the sum of the colloid force balance (\ref{eqn:FBC}) and the solvent force balance (\ref{eqn:FBS}), yielding in the planar and flow-free geometry of interest
\begin{equation*}
 \label{eqn:FBtot}
  f_z^e(z) - \partial_z P_{\text{tot}}(z) = 0,
 \numberthis
\end{equation*}
\noindent where $P_{\text{tot}}(z) \equiv P(z) + P_s(z)$. From the total force balance (\ref{eqn:FBtot}),
the osmotic pressure $\Pi \equiv \int_{z_{\text{res}}}^{z_b} \mathrm{d}z f_z^e(z)
$, defined as the magnitude of the force per unit area exerted on the membrane, follows as 
$P_{\text{tot}}(z_b) - P_s(z_{\text{res}}).$
As the total bulk pressure decomposes into colloid and solvent contributions as
$P_{\text{tot}}(z_b) 
  = 
  \rho k_BT + P_s(z_b)$,
the osmotic pressure reads
\begin{equation*}
 \label{eqn:ActiveHoff}
  \Pi 
  =
  \rho k_BT + \Delta P_s. 
 \numberthis
\end{equation*}
In equilibrium, Eq. (\ref{eqn:ActiveHoff}) reduces to Van't Hoff's result $\Pi = \rho k_BT$ on account of $\Delta P_s = 0$. Activity increases the osmotic pressure by increasing the solvent pressure with respect to the reservoir by $\Delta P_s$, which is the key result of this work. Together, Eq. (\ref{eqn:ActiveHoff}) and (\ref{eqn:DeltaPs}) generalize Van't Hoff's law to active suspensions.\\
\begin{figure}
 \includegraphics[width=\linewidth]{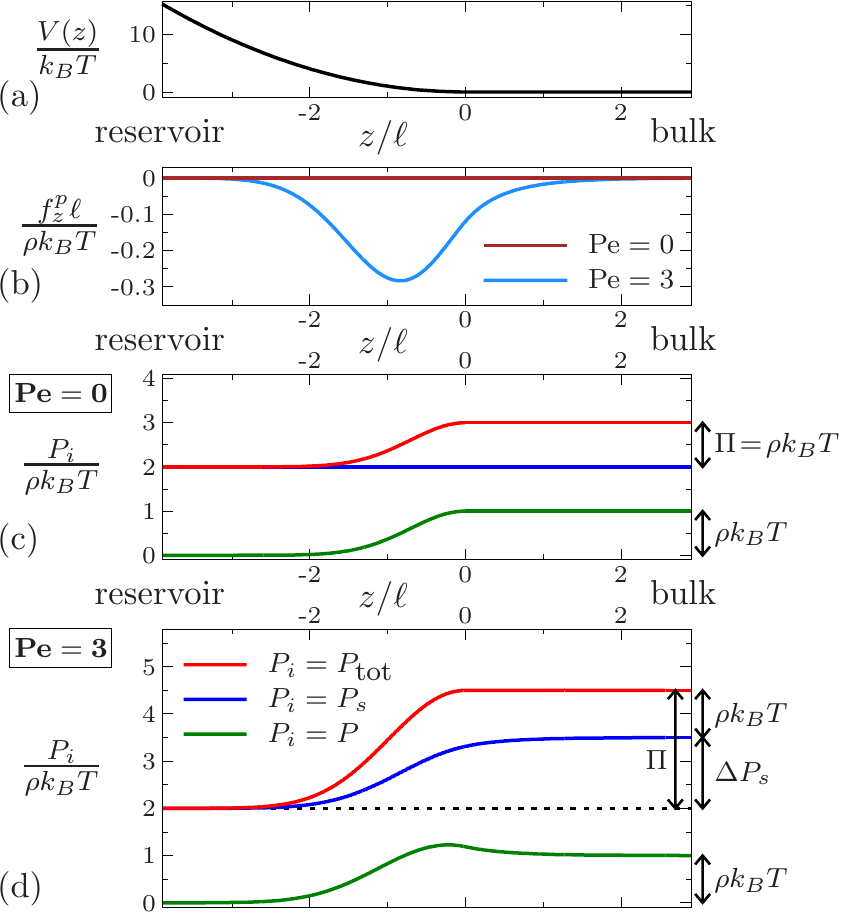}
 \caption{(a) External potential $V(z)$ modelling the planar membrane of Fig. \ref{fig:NicePicture} that separates the reservoir at $z_{\text{res}} = -4 \ell$ from the bulk suspension at $z_b = 3 \ell$. (b) The steady state propulsion body force $f^p_z(z)$, at activity Pe. Passive (c) and active (d) pressure profiles of the colloids $P(z)$, the solvent $P_s(z)$, and the total suspension $P_{\text{tot}}(z) = P(z) + P_s(z)$. For the passive system (Pe $=0$), $P_s(z)$ is constant, such that the osmotic pressure $\Pi = P_{\text{tot}}(z_b) - P_{\text{tot}}(z_{\text{res}})$ equals the bulk colloid pressure $P(z_b) = \rho k_BT$. For the active system (Pe $=3$), the reaction body force $-f^p_z(z)$ increases the bulk solvent pressure $P_s(z_b)$, as well as $P_{\text{tot}}(z_b)$ and $\Pi$, by $\Delta P_s$.}
 \label{fig:IsoHarm}
\end{figure}	
	\indent To clarify these concepts further, we have solved the Smoluchowski equation (\ref{eqn:FP}) numerically for a system of spheres subject to a propulsion force, characterized by Peclet number Pe $\equiv (\gamma_t\gamma_r)^{\frac{1}{2}}v_0/k_BT$, in the planar geometry modelling the setting of Fig. \ref{fig:NicePicture}. The membrane, felt by the colloids only, is modelled by the soft potential $V(z) = \lambda k_BT (z/\ell)^2$ for $z < 0$ and $V(z) = 0$ for $z \geq 0$ (Fig. \ref{fig:IsoHarm}(a)), i.e. there is no torque. Here $\lambda = 1$ is the strength of the potential, and $\ell \equiv (\gamma_r/\gamma_t)^{\frac{1}{2}}$ is the appropriate unit of length, which is of the order of the (effective) particle size upon using Stokes relations for $\gamma_t$ and $\gamma_r$. Fig. \ref{fig:IsoHarm}(b) shows the profile of the propulsion body force $f^p_z(z)$. Whereas $f^p_z(z) = 0$ for a passive system (Pe $=0$), an active system displays a nonzero polarization $m_z(z)$, and thus by Eq. (\ref{eqn:fp}) a propulsion body force $f^p_z(z)$, in the vicinity of the membrane directed towards the membrane. This well-known effect \cite{WallHugging1, WallHugging2, WallHugging3,GompperWallAccumulation,Lee,HaganWallAccumulation} is in this case caused by colloids persistently propelling `into' the repulsive membrane. Fig. \ref{fig:IsoHarm}(c), for Pe$=0$, shows the pressure profiles $P(z)$ of the passive colloids, $P_s(z)$ of the solvent, and $P_{\text{tot}}(z)$ of the total passive suspension. Here the reaction body force $-f^p_z(z) = 0$, and hence the solvent pressure $P_s(z)$ is constant, as argued before. It is only due to the bulk colloid pressure $P(z_b) = \rho k_BT$ that the total bulk pressure $P_{\text{tot}}(z_b)$ is higher than the total reservoir pressure $P_{\text{tot}}(z_{\text{res}})$. The osmotic pressure $\Pi = P_{\text{tot}}(z_b) - P_{\text{tot}}(z_{\text{res}})$ is therefore equal to $\rho k_BT$. The profiles for an active system (Pe $= 3$), displayed in Fig. \ref{fig:IsoHarm}(d), show that the solvent bulk pressure $P_s(z_b)$ exceeds the solvent reservoir pressure $P_s(z_{\text{res}})$. This is caused by the reaction body force $-f^p_z(z)$, that pushes solvent towards the bulk, as pictured in Fig. \ref{fig:NicePicture}(b). As a result, both the total bulk pressure $P_{\text{tot}}(z_b)$ and the osmotic pressure $\Pi$ exceed their passive counterparts by $\Delta P_s = \gamma_t \gamma_r v_0^2\rho/6k_BT$ on account of Eq. (\ref{eqn:DeltaPs}) for the torque-free potential of interest here.\\
\begin{figure}[t]
 \includegraphics[width=\linewidth]{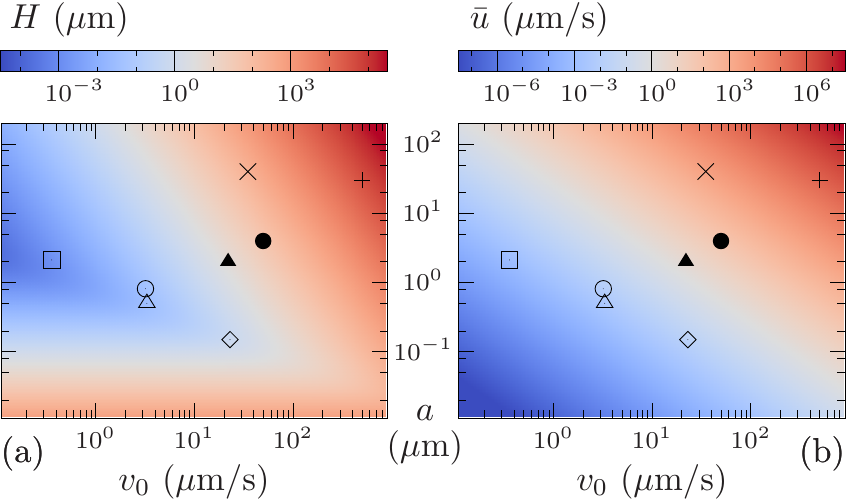}
 \caption{(a) Predicted rise $H = H_0 + \Delta H$ in Fig. \ref{fig:NicePicture}, for spherical particles of radius $a$ at propulsion speed $v_0$ at packing fraction $0.01$ in water. (b) Predicted mean solvent velocity $\bar{u}$ in Fig. \ref{fig:PipeFlow}, for a cylindrical pipe of radius $5a$ and length $100a$. Symbols denote literature values of ($v_0$,$a$) combinations of experimentally realized self-propelled colloids $+$ \cite{Pepijn}, $\times$ \cite{Vicario}, $\diamond$ \cite{Wilson}, $\circ$ \cite{Howse}, $\triangle$ \cite{Palacci2010}, $\square$ \cite{Volpe}; and motile bacteria \textbullet \ \cite{Mussler}, $\blacktriangle$ \cite{Drescher}.}
 \label{fig:Experiment}
\end{figure}
\section*{Experimental predictions}
\noindent The experiments that have adressed the pressure of active systems \cite{EoSExperiment,BradySupernova,Junot2017} are few in number. In particular, the osmotic pressure has never been measured directly. Despite the simplicity of the ABP model, that neglects e.g. hydrodynamic interactions, our expression for the osmotic pressure does allow to estimate the order of magnitude of the meniscus height difference $H = \Pi / (\rho^m_{s} g)$ that is to be expected in the experiment sketched in Fig. \ref{fig:NicePicture}. Here we focus on an aqueous dispersion (mass density $\rho^m_s = 1 \ \text{kg}/\text{dm}^3$) of active hard spheres of radius $a$, with friction coefficients given by the Stokes relations $\gamma_t = 6\pi\eta a$ and $\gamma_r = 8 \pi \eta a^3$, subject to Earth's gravitational acceleration $g$, at room temperature, and at packing fraction $0.01$ that should mimic the ideal (non-interaction) conditions.  The predicted height differences $H$ are shown in Fig. \ref{fig:Experiment}(a). Whereas the passive osmotic pressure $\rho k_B T$ induces a passive rise $H_0 \sim a^{-3}$ too small to measure for colloidal particles, activity induces an additional rise $\Delta H \sim \Delta P_s \sim a v_0^2$ that brings $H = H_0 + \Delta H$ up to the regime of micrometers \cite{Vicario,Mussler,Drescher} or even millimeters \cite{Pepijn} for the larger values of propulsion speed $v_0$ and particle size $a$ of experimentally realized microswimmers.\\
	\indent To determine experimentally that the activity-induced increase in osmotic pressure results from the increase in \emph{solvent} pressure $\Delta P_s$, we propose to confine active particles by a membrane to one half of an open, horizontal pipe, for which gravity plays no role, as illustrated in Fig. \ref{fig:PipeFlow}. Applying \emph{equal} solvent pressures to either side of the pipe, rather than the no-flux boundary condition before, results in a steady state where the reaction body force near the membrane $-\mb{f}^p$ drives a steady solvent flow $\mb{u}(\mb{r})$ through the pipe (as seen in the lab frame), according to Eq. (\ref{eqn:FBS}). In the limit $|\mb{u}(\mb{r})| \ll v_0$, and for a cylindrical pipe, this flow velocity is identical to the Poisseuile flow that would be generated in a pipe filled with only solvent upon applying the solvent pressure difference $\Delta P_s$ of Eq. (\ref{eqn:DeltaPs}) between either end of the pipe. For a derivation see section 6 of the ESI$^{\dag}$. The predicted mean solvent velocity $\bar{u} \sim a^2 v_0^2$ is shown in Fig. \ref{fig:Experiment}(b) as a function of the propulsion speed $v_0$ and colloid radius $a$, for a pipe of radius $5a$ and length $100a$. For the larger values of $v_0$ and $a$ of experimentally realized swimmers \cite{Drescher, Mussler, Vicario, Pepijn}, the solvent velocity $\bar{u}$ (although not satisfying $\bar{u} \ll v_0$ in all cases) is predicted to be on the order of micrometers per second, and hence easily detectable, e.g. by using tracer particles.\\
\section*{The solvent chemical potential}
\noindent We now return to the the original setting of Fig. \ref{fig:NicePicture} to interpret the solvent pressure difference $\Delta P_s$ between the suspension and the reservoir. Even though the active suspension is out of equilibrium, the solvent pressure $P_s(z)$ can still be used to define a meaningful intrinsic solvent chemical potential $\mu_s^{\text{int}}(z)$ by the (Gibbs-Duhem like) relation
  $\rho_s(z) \partial_z \mu_s^{\text{int}}(z) 
  =
  \partial_z P_s(z)$,
\noindent with $\rho_s(z)$ the number density of the solvent 
(see section 3 of the ESI$^{\dag}$ for details). Hence, the solvent pressure difference $\Delta P_s$ is accompanied by a difference in the intrinsic solvent chemical potential
\begin{equation*}
 \label{eqn:DeltaMus}
  \Delta \mu_s \equiv \mu_s^{\text{int}}(z_b) - \mu_s^{\text{int}}(z_{\text{res}})
  = \int_{z_{\text{res}}}^{z_b} \mathrm{d}z \frac{\partial_z P_s(z)}{\rho_s(z)}.
 \numberthis
\end{equation*}
\noindent We can thus rephrase our findings as follows. Activity increases the solvent chemical potential of the bulk suspension from the reservoir value $\mu_s$ to $\mu_s + \Delta \mu_s$. The total bulk pressure $P_{\text{tot}}(\rho,\mu_s+\Delta \mu_s) = \rho k_B T + P_s(\mu_s+\Delta \mu_s)$ increases accordingly, such that the osmotic pressure $\Pi = P_{\text{tot}}(\rho,\mu_s + \Delta \mu_s) - P_s(\mu_s)$, which is the difference between the total bulk pressure and the reservoir pressure, now equals $\Pi = \rho k_BT + \Delta P_s$, where $\Delta P_s = P_s(\mu_s + \Delta \mu_s) - P_s (\mu_s)$ is the difference in solvent pressures accompanying the difference in solvent chemical potentials.\\
\begin{figure}[b]
\center
 \includegraphics[width=.8\linewidth]{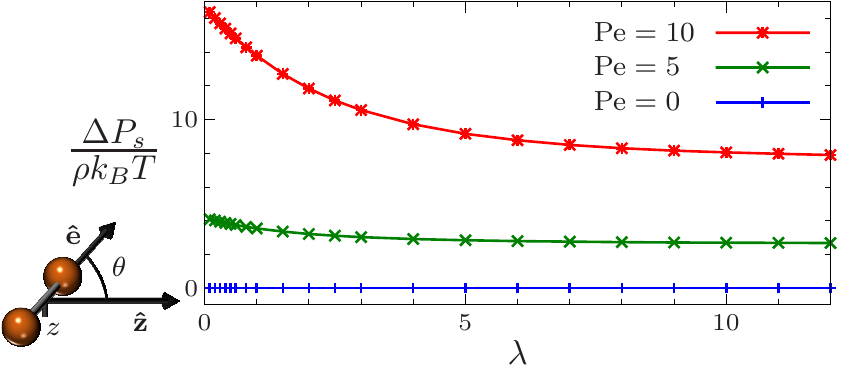}
 \caption{Increase in bulk solvent pressure $\Delta P_s$ as a function of the strength $\lambda$ of the soft colloid-membrane interaction potential in the setting of Fig. \ref{fig:NicePicture} for active dumbells at varying activity Pe. For active systems (Pe$>0$) $\Delta P_s$ depends on $\lambda$.}
 \label{fig:Osmo2}
\end{figure}
	\indent In this light, we address the second term of Eq. (\ref{eqn:DeltaPs}), present for anisotropic colloids experiencing a torque $-\partial_{\theta}V(z,\theta)$. To investigate the implications of this term, we have solved the Smoluchowski equation (\ref{eqn:FP}) for active dumbells, consisting of two point particles with separation $\ell = (\gamma_r/\gamma_t)^{\frac{1}{2}}$. Both point particles are subject to the same membrane potential $V(z) = \lambda k_B T (z/\ell)^2$ for $z<0$ as before, where the strength parameter $\lambda$ can now be varied. The resulting potential acting on a dumbell, $V(z,\theta) = V(z+\frac{\ell}{2}\cos\theta) + V(z-\frac{\ell}{2}\cos\theta)$, exerts a nonzero torque $-\partial_{\theta}V(z,\theta)$, that tends to align dumbells parallel to the wall. Fig. \ref{fig:Osmo2} shows the resulting increase in solvent pressure $\Delta P_s$, calculated from Eq. (\ref{eqn:DeltaPs}), for different activities Pe as a function of the strength $\lambda$ of the colloid-membrane interaction potential. For Pe $> 0$, $\Delta P_s$ decreases with $\lambda$. The reason for this decrease is that the torque generated by the potential rotates the particles that propel `into' the membrane, and thereby influences the shape of the polarization profile $m_z(z)$. As it turns out, the torque reduces the total polarization near the membrane $-\int_{z_{\text{res}}}^{z_b}\mathrm{d}z m_z(z)$, and by that also the integrated reaction body force $-\int_{z_{\text{res}}}^{z_b}\mathrm{d}zf_z^p(z)$ that pushes solvent towards the suspension, see Eq. (\ref{eqn:fp}). Consequently, the increase in solvent pressure $\Delta P_s$ decreases as the strength of the colloid-membrane interaction potential increases. The same dependence was found in Ref. \cite{Cates} for ellipsoidal particles under the assumptions that the distribution $\psi(z,\theta)$ attains its bulk value already at $z=0$, and that the effect of ellipses that only feel the potential partially is negligible.
We thus confirm the conclusion of Ref. \cite{Cates} that the second term of Eq. (\ref{eqn:DeltaPs}) depends on the precise form of the colloid-membrane interaction potential $V(z,\theta)$, by a numerical solution $\psi(z,\theta)$ that does not require any further assumptions.\\
	\indent Whereas in Ref. \cite{Cates} this finding was reason to question whether the osmotic pressure is a state function, we emphasize it is the bulk state of the suspension itself that depends on the colloid-membrane potential. To appreciate its consequences, we note that in equilibrium the ensemble of reservoir \emph{and} suspension is specified by the state variables $(\mu_s,\rho,T)$, since the solvent chemical potential of the reservoir $\mu_s$ sets the same chemical potential in the suspension. The fact that for an active system the solvent pressure difference $\Delta P_s$ - and thereby also the chemical potential difference $\Delta \mu_s$ - generally depends on the colloid-membrane interaction potential, implies that a complete specification of the ensemble requires an additional state variable, e.g. the bulk solvent chemical potential $\mu_s^b \equiv \mu_s+\Delta \mu_s$. In fact, upon including effective colloid-colloid interactions, the activity is also required as a state variable, e.g. in terms of $v_0$ (see section 3 of the ESI$^{\dag}$). A complete set of (intensive) state variables therefore reads $(\mu_s,\mu_s^b,\rho,T,v_0)$. All the mentioned pressures, including the osmotic pressure, \emph{are} state functions of these variables.\\

\section*{Conclusions}
\noindent We have generalized Van't Hoff's law to active suspensions. We have shown that the active particles exert a net reaction force on the solvent, an effect that we predict to be experimentally measurable either as a solvent flow through a semipermeable membrane confining the active suspension to one side of an open pipe, or as a macroscopic rise of the suspension meniscus in a U-pipe experiment. In the latter case, the reaction force increases the solvent pressure of the suspension, and thereby the solvent chemical potential. Remarkably, this increase, and thereby the bulk state of the suspension \emph{itself}, depends on the details of the colloid-membrane interactions. The osmotic pressure \emph{is} a state function of (amongst others) the solvent chemical potential; it \emph{does} depend on the details of the colloid-membrane interactions, but only \emph{via} the solvent chemical potential.

\section*{Discussion}
\noindent The predictions of Eq. (\ref{eqn:DeltaPs}) and of Fig. \ref{fig:Experiment} are made for active particles whose orientation changes only by rotational diffusion with a rate that follows from the Stokes-Einstein relations for spherical particles. The corresponding typical reorientation time $\tau_r$, equal to $\gamma_r/k_BT$ in this case, is shown by Eq. (\ref{eqn:DeltaPs}) to be proportional to the excess solvent pressure $\Delta P_s$. In fact, the result $\Delta P_s \sim \tau_r$ is more general \cite{BradySwimPressure}, because the typical time $\tau_r$ that a particle spends propelling `into' the membrane determines the magnitude of the time-averaged reaction force it exerts on the solvent, and thus of the excess solvent pressure $\Delta P_s$. In general, this reorientation time $\tau_r$ depends on more factors, for instance on the details of the propulsion mechanism of the active particle, and on its (hydrodynamic) interaction with the membrane \cite{Lauga2006,WallHugging2}. 
\noindent An interesting example of the latter type occurs for the square-shaped particles simulated in Ref. \cite{VassilisSquares}. These particles tend to form a crystal phase next to the membrane, with the majority of particles facing the membrane \cite{VassilisPrivateCommunication}. This effect increases $\tau_r$, and thereby the excess solvent pressure $\Delta P_s$, dramatically.\\

\noindent In the ESI$^{\dag}$, we generalize the framework presented here to include interactions. The active version of Van't Hoff's law  (\ref{eqn:ActiveHoff}) then generalizes to $\Pi = P(\rho,\mu_s^b,v_0)+\Delta P_s$, where $P(\rho,\mu_s^b,v_0)$ now denotes the full pressure (ideal gas plus virial contributions) of the effective colloids-only system that is characterized by $(\rho,\mu_s^b,v_0)$. Hence, the functional form of the osmotic pressure $\Pi$ differs from its passive expression only by the excess solvent pressure $\Delta P_s$. This excess pressure $\Delta P_s$ again depends on the membrane potential, except in the absence of any torque interactions between either the particles and the membrane, or between the particles themselves. In the absence of such torques, $\Delta P_s$ again reduces to the known swim pressure. While this swim pressure is a linear function of $\rho$ at low colloid densities, cf. Eq. (\ref{eqn:DeltaPs}), it typically becomes a decreasing function of $\rho$ at high densities \cite{MarchettiSwimPressure,CatesInteractingSpheres,BradyReview,Vassilis2016}. For general interactions, it remains true that the difference in solvent pressure $\Delta P_s$ is accompanied by a difference in solvent chemical potential $\Delta \mu_s$, and that the osmotic pressure \emph{is} a state function of the variables $(\mu_s,\mu_s^b,\rho,T,v_0)$.\\

\noindent Crucial in our approach is that activity enters the colloid force balance (\ref{eqn:FBC}) as the body force $\mb{f}^p(\mb{r})$, cf. Ref. \cite{BradyBodyForce}, whereas the local pressure $P(\mb{r})=\rho(\mb{r})k_BT$ is of the same form as in equilibrium. Our approach follows Speck and Jack \cite{SpeckJack}, who showed that the bulk colloid pressure $\rho k_BT$ represents momentum flux of non-interacting colloids. In the interacting case, the local pressure tensor $P(\mb{r})$ generalizes to a local pressure tensor, that consists of both momentum flux and a term accounting for interaction forces (see section 1 of the ESI$^{\dag}$), as the conventional local pressure tensor does \cite{RowlinsonWidom}. Representing activity by a body force contrasts the approach of some authors who account for activity, in a colloid-only picture, by modifying the local pressure \cite{Lowen,MarconiMaggi,StarkPhaseSeparation,Berend2016}. Whereas our approach is valid for general particle-particle and particle-wall interactions$^{\dag}$, this `activity-modified' local pressure has only been defined for isotropic particles, and indeed its derivation \cite{Falasco2016,Fily2017} does not seem to be extendable to systems with torque interactions.\\
	\indent The generalizations of the force balances (\ref{eqn:FBC}), (\ref{eqn:FBS}) and (\ref{eqn:FBtot}) to an interacting suspension (see section 3 of the ESI$^{\dag}$) can readily be applied to other typical phenomena exhibited by active systems, such as the motility-induced phase separation (MIPS) of purely repulsive particles \cite{ExpMIPS,MIPSReview,MarchettiABPReview,Stark2014,Fielding2014}. Strikingly, the interface of the phase coexistence generated by MIPS was found to have a negative interfacial tension, defined in the colloid-only picture in terms of the activity-modified pressure tensor \cite{Lowen}. This begs the question what this negative interfacial tension - and its interpretation \cite{Lowen, SpeckStochastic} - translate into in the picture presented here, both in the colloid-only sense and upon taking the solvent into account.\\
	\indent We foresee the picture presented here to form a basis for making headway in understanding this extraordinary world of active matter physics, and express the hope that the predictions of Fig. \ref{fig:Experiment} will stimulate experimental efforts to actually measure the osmotic pressure $\Pi$ and the associated solvent pressure difference $\Delta P_s$ of active suspensions.

\section*{Acknowledgments}
We acknowledge Bob Evans and Bram Bet for helpful discussions. This work is part of the D-ITP consortium, a program of the Netherlands Organisation for Scientific Research (NWO) that is funded by the Dutch Ministry of Education, Culture and Science (OCW).



\putbib[../../../../Literature/Database/ActiveMatter/ActiveMatter]
\end{bibunit}


\renewcommand*{\thefootnote}{\arabic{footnote}} 

\begin{bibunit}[rsc]

\onecolumngrid
\clearpage

\vspace{2.0cm}
\section*{\Large Electronic Supplementary Information (ESI)}
\vspace{0.7cm}
\section{1. Smoluchowski equation and momentum balance for interacting ABPs}

\label{IntMB}
\noindent In this section we derive the Smoluchowski equation and the momentum balance for interacting ABPs, and show that they reduce in the non-interacting case to Eq. (\ref{eqn:FP}) and Eq. (\ref{eqn:FBC}) of the main text, respectively. The local pressure $P(\mb{r}) = \rho(\mb{r}) k_BT$ of the main text shall thereby be identified as momentum flux of the colloids.\\
	\indent In order to establish a well-defined momentum balance for the overdamped ABPs, we start by considering dynamics \emph{with} inertia. Every particle $i\in \{1, \dots ,N\}$ is then characterized by its center-of-mass position $\mb{r}_i(t)$ and its orientation $\mb{\hat{e}}_i(t)$, as well as by its translational velocity $\mb{v}_i(t) = \mb{\dot{r}}_i(t)$ and its angular velocity $\mbs{\omega}_i(t)$ satisfying $\mb{\dot{\hat{e}}}_i = \mbs{\omega}_i \times \mb{\hat{e}}_i$. The time evolution is governed by the Langevin equations
\begin{flalign*}
 \label{eqn:Langevin}
    m \mb{\dot{v}}_i &=  
      - \gamma_t \mb{v}_i
      - \mbs{\nabla}_i\left[V(\mb{r}_i,\mb{\hat{e}}_i) +U^{\text{eff}}(\mb{r}^N,\mb{\hat{e}}^N)\right]
      + \gamma_t v_0 \mb{\hat{e}}_i 
      +  \sqrt{2\gamma_tk_BT}\mbs{\eta}^t_i(t)
  \quad \text{and}
     \\
    I \mbs{\dot{\omega}}_i &= 
      - \gamma_r \mbs{\omega}_i 
      - \rot_i \left[V(\mb{r}_i,\mb{\hat{e}}_i) +U^{\text{eff}}(\mb{r}^N,\mb{\hat{e}}^N)\right]
      + \sqrt{2\gamma_rk_BT}\mbs{\eta}^r_i(t).
 \numberthis&&
\end{flalign*}
\noindent Here $m$ is the mass of a particle, $I$ its moment of inertia, and $\rot_i \equiv \mb{\hat{e}}_i \times \mbs{\nabla}_{\mb{\hat{e}}_i}$ denotes the rotation operator. Eq. (\ref{eqn:Langevin}) expresses every particle experiencing i) a linear frictional force and torque due to the solvent \footnote{For anisotropic particles, $\gamma_t$ should actually be replaced by an orientation-dependent friction matrix. This does not qualitatively change our results.}, ii) forces/torques due to an external potential $V(\mb{r},\mb{\hat{e}})$ and due to an effective interaction potential $U^{\text{eff}}(\mb{r}^N,\mb{\hat{e}}^N)$, that captures the effective interactions between the colloids in the solvent (we neglect hydrodynamic interactions), iii) a self-propulsion force with magnitude $\gamma_tv_0$ in direction $\mb{\hat{e}}_i$, and iv) Brownian forces/torques that are governed by the white Gaussian noises $\mbs{\eta}^t_i(t)$ and $\mbs{\eta}^r_i(t)$, satisfying $\langle \eta^t_{i,\alpha}(t) \eta^t_{j,\beta}(t') \rangle = \delta _{ij} \delta_{\alpha\beta} \delta(t-t')$ and $\langle \eta^r_{i,\alpha}(t) \eta^r_{j,\beta}(t') \rangle = \delta_{ij} \delta_{\alpha\beta} \delta(t-t')$, respectively, for $i,j \in \{1,\dots,N\}$ and $\alpha, \beta \in\{x,y,z\}$. The notation $\mb{r}^N$ is shorthand for $\{\mb{r}_1, \dots , \mb{r}_N\}$, and similarly for $\mb{\hat{e}}^N$ etc. One can check from e.g. \cite{Palacci2010} that the time scale $m/\gamma_t$ for changes in the velocity $\mb{v}$ is much smaller than the time scale $v_0^{-1} (\gamma_r/\gamma_t)^{1/2}$ for changes in the position $\mb{r}$. Similarly, the angular velocity changes on a time scale $I/\gamma_r$, which is much smaller than the rotational time scale $\gamma_r/k_BT$. This is the essence of overdamped motion. The overdamped equations of motion are obtained by taking the limit $m,I \rightarrow 0$ in Eq. (\ref{eqn:Langevin}), but for now we consider finite $m,I$. The probability distribution of the noise terms in Eq. (\ref{eqn:Langevin}) induces the probability distribution function $f^{(N)}(\mb{r}^N,\mb{e}^N,\mb{v}^N,\mbs{\omega}^N,t)$ at time $t$ to evolve according to the Fokker-Planck equation \cite{vanKampen}
\begin{flalign*}
 \label{eqn:InertFP}
  &\partial_t f^{(N)} = \numberthis \\
  &- \sum_i \mbs{\nabla}_i \cdot \left(f^{(N)}\mb{v}_i\right)
    - \frac{\gamma_t}{m} \sum_i \mbs{\nabla}_{\mb{v}_i} \cdot
      \bigg\{ 
       \left(
        - \mb{v}_i
        - \gamma_t^{-1} \mbs{\nabla}_i \left[V(\mb{r}_i,\mb{\hat{e}}_i)
                                         +U^{\text{eff}}(\mb{r}^N,\mb{\hat{e}}^N)
                                   \right]
        + v_0 \mb{\hat{e}}_i 
        - \frac{k_BT}{m}\mbs{\nabla}_{\mb{v}_i}
       \right)
      f^{(N)}
      \bigg\} \\
  &- \sum_i \rot_i \cdot \left(f^{(N)} \mbs{\omega}_i\right)
   -  \frac{\gamma_r}{I} \sum_i\mbs{\nabla}_{\mbs{\omega}_i} \cdot
      \bigg\{ 
       \left(
        - \mbs{\omega}_i
        - \gamma_r^{-1} \rot_i \left[V(\mb{r}_i,\mb{\hat{e}}_i)
                                         +U^{\text{eff}}(\mb{r}^N,\mb{\hat{e}}^N)
                                   \right]
        - \frac{k_BT}{I}\mbs{\nabla}_{\mbs{\omega}_i}
       \right)
      f^{(N)}
      \bigg\}.
\end{flalign*}
\noindent In order to integrate out the velocity degrees of freedom we define the relevant distribution functions
\begin{flalign*}
 \label{eqn:Defs}
 \begin{alignedat}{1}
  \psi^{(N)}(\mb{r}^N,\mb{\hat{e}}^N,t) 
  &\equiv \int \mathrm{d}\mb{v}^N\mathrm{d}\mbs{\omega}^N 
        f^{(N)}(\mb{r}^N,\mb{\hat{e}}^N,\mb{v}^N,\mbs{\omega}^N,t),\\
  \psi^{(N)}(\mb{r}^N,\mb{\hat{e}}^N,t) \mb{\bar{v}}_i(\mb{r}^N,\mb{\hat{e}}^N,t)
  &\equiv 
   \int \mathrm{d}\mb{v}^N\mathrm{d}\mbs{\omega}^N 
     f^{(N)}(\mb{r}^N,\mb{\hat{e}}^N,\mb{v}^N,\mbs{\omega}^N,t) \mb{v}_i,\\
  \psi^{(N)}(\mb{r}^N,\mb{\hat{e}}^N,t) \mbs{\bar{\omega}}_i(\mb{r}^N,\mb{\hat{e}}^N,t)
  &\equiv 
   \int \mathrm{d}\mb{v}^N\mathrm{d}\mbs{\omega}^N 
     f^{(N)}(\mb{r}^N,\mb{\hat{e}}^N,\mb{v}^N,\mbs{\omega}^N,t) \mbs{\omega}_i,
 \end{alignedat}
 \numberthis &&
\end{flalign*}
\noindent such that the zeroth moment of Eq. (\ref{eqn:InertFP}) reads
\begin{flalign*}
 \label{eqn:M0}
  \partial_t\psi^{(N)}
   =
  -\sum_i \mbs{\nabla}_i \cdot \left(\psi^{(N)} \mb{\bar{v}}_i)\right)
  -\sum_i \rot_i \cdot \left(\psi^{(N)} \mbs{\bar{\omega}}_i)\right);
 \numberthis &&
\end{flalign*}
\noindent its first moment in $\mb{v}_i$ reads
\begin{flalign*}
 \label{eqn:M1v}
  \partial_t \left(\psi^{(N)} \mb{\bar{v}}_i \right)
   =
  &- \sum_j \mbs{\nabla}_j \cdot 
     \left( \int\mathrm{d}\mb{v}^N\mathrm{d}\mbs{\omega}^Nf^{(N)}\mb{v}_j \otimes \mb{v}_i \right)
  - \sum_j \rot_j \cdot 
     \left( \int\mathrm{d}\mb{v}^N\mathrm{d}\mbs{\omega}^Nf^{(N)}\mbs{\omega}_j \otimes \mb{v}_i \right)\\
  &+ \frac{\gamma_t}{m}
    \Big( -\mb{\bar{v}_i} 
           - \gamma_t^{-1} \mbs{\nabla}_i\left[V(\mb{r}_i,\mb{\hat{e}}_i) 
                                           +U^{\text{eff}}(\mb{r}^N,\mb{\hat{e}}^N)\right]
           +v_0\mb{\hat{e}}_i
    \Big)\psi^{(N)};
 \numberthis &&
\end{flalign*}
\noindent and its first moment in $\mbs{\omega}_i$ reads
\begin{flalign*}
 \label{eqn:M1Omega}  
  \partial_t \left(\psi^{(N)} \mbs{\bar{\omega}}_i \right)
   =
  &- \sum_j \mbs{\nabla}_j \cdot 
     \left( \int\mathrm{d}\mb{v}^N\mathrm{d}\mbs{\omega}^Nf^{(N)}\mb{v}_j \otimes \mbs{\omega}_i \right)
  - \sum_j \rot_j \cdot 
     \left( \int\mathrm{d}\mb{v}^N\mathrm{d}\mbs{\omega}^Nf^{(N)}\mbs{\omega}_j\otimes\mbs{\omega}_i\right)\\
  &+ \frac{\gamma_r}{I}
    \Big( -\mbs{\bar{\omega}_i} 
           - \gamma_r^{-1} \rot_i\left[V(\mb{r}_i,\mb{\hat{e}}_i) 
                                           +U^{\text{eff}}(\mb{r}^N,\mb{\hat{e}}^N)\right]
    \Big)\psi^{(N)}.
 \numberthis &&
\end{flalign*}
Here we used the notation $\mb{a} \otimes \mb{b}$ to denote the dyadic product of two vectors $\mb{a}$ and $\mb{b}$. Next, we integrate out the remaining degrees of freedom of all but one particles. At this point we assume the effective interactions to be pairwise, i.e. $U^{\text{eff}}(\mb{r}^N,\mb{\hat{e}}^N) = \sum_{i<j} \phi^{\text{eff}}_{\mb{\hat{e}}_i\mb{\hat{e}}_j}(\mb{r}_j-\mb{r}_i)$. This approximation is made purely for simplification purposes; the methods presented here can be extended to three- and higher-body interactions \cite{Archer2008,RowlinsonWidom}. We define the one-body distribution functions
\begin{flalign*}
 \label{eqn:OneBodyDefs}
  f(\mb{r}_1,\mb{\hat{e}}_1,\mb{v}_1,\mbs{\omega}_1,t)
   &\equiv
  N \int \mathrm{d}\mb{r}^{(2\rightarrow N)} \mathrm{d}\mb{\hat{e}}^{(2\rightarrow N)}
         \mathrm{d}\mb{v}^{(2\rightarrow N)} \mathrm{d}\mbs{\omega}^{(2\rightarrow N)}
           f^{(N)}(\mb{r}^N,\mb{\hat{e}}^N,\mb{v}^N,\mbs{\omega}^N,t), \\
  \psi(\mb{r}_1,\mb{\hat{e}}_1,t)
   &\equiv
  N \int \mathrm{d}\mb{r}^{(2\rightarrow N)} \mathrm{d}\mb{\hat{e}}^{(2\rightarrow N)}
           \psi^{(N)}(\mb{r}^N,\mb{\hat{e}}^N,\mb{v}^N,\mbs{\omega}^N,t), \\  
  \psi(\mb{r}_1,\mb{\hat{e}}_1,t) \mb{\bar{v}}(\mb{r}_1,\mb{\hat{e}}_1,t)
   &\equiv
  N \int \mathrm{d}\mb{r}^{(2\rightarrow N)} \mathrm{d}\mb{\hat{e}}^{(2\rightarrow N)}
           \psi^{(N)}(\mb{r}^N,\mb{\hat{e}}^N,t)
           \mb{\bar{v}}_1(\mb{r}^N,\mb{\hat{e}}^N,t), \\  
  \psi(\mb{r}_1,\mb{\hat{e}}_1,t) \mbs{\bar{\omega}}(\mb{r}_1,\mb{\hat{e}}_1,t)
   &\equiv
  N \int \mathrm{d}\mb{r}^{(2\rightarrow N)} \mathrm{d}\mb{\hat{e}}^{(2\rightarrow N)}
           \psi^{(N)}(\mb{r}^N,\mb{\hat{e}}^N,t)
           \mbs{\bar{\omega}}_1(\mb{r}^N,\mb{\hat{e}}^N,t),
 \numberthis &&
\end{flalign*}
\noindent and the two-body distribution function
\begin{flalign*}
 \label{eqn:TwoBodyDef}
  \psi^{(2)}_{\mb{\hat{e}}_1\mb{\hat{e}}_2}(\mb{r}_1,\mb{r}_2,t)
   \equiv
  N(N-1) \int \mathrm{d}\mb{r}^{(3\rightarrow N)} \mathrm{d}\mb{\hat{e}}^{(3\rightarrow N)}
           \psi^{(N)}(\mb{r}^N,\mb{\hat{e}}^N,\mb{v}^N,\mbs{\omega}^N,t),
 \numberthis &&
\end{flalign*}
\noindent where $\int\mathrm{d}\mb{r}^{(n\rightarrow N)}$ denotes an integration over $\mb{r}_n,\mb{r}_{n+1},\dots,\mb{r}_N$ (and similarly for $\int\mathrm{d}\mb{\hat{e}}^{(n\rightarrow N)}$ etc). Integrating over all but one particles then yields for Eq. (\ref{eqn:M0})
\begin{flalign*}
 \label{eqn:OneBodyM0}
  \partial_t \psi   =
  -\mbs{\nabla}_1 \cdot \left(\psi \mb{\bar{v}}\right)
  -\rot_1 \cdot \left(\psi\mbs{\bar{\omega}}\right);
 \numberthis &&
\end{flalign*}
\noindent for Eq. (\ref{eqn:M1v})
\begin{flalign*}
 \label{eqn:OneBodyM1v}
  \partial_t \left(\psi \mb{\bar{v}} \right)
   = 
  &- \mbs{\nabla}_1 \cdot 
     \left(\int\mathrm{d}\mb{v}_1\mathrm{d}\mbs{\omega}_1
            f \mb{v}_1 \otimes \mb{v}_1 
     \right)
  - \rot_1 \cdot 
     \left(\int\mathrm{d}\mb{v}_1\mathrm{d}\mbs{\omega}_1
            f \mbs{\omega}_1 \otimes \mb{v}_1 
     \right)\numberthis \\
  &+ \frac{\gamma_t}{m}
     \Big( \left[-\mb{\bar{v}} - 
            \gamma_t^{-1} \mbs{\nabla}_1 V(\mb{r}_1,\mb{\hat{e}}_1) 
            + v_0 \mb{\hat{e}}_1
           \right]\psi
           -\gamma_t^{-1} 
             \int\mathrm{d}\mb{r}_2\mathrm{d}\mb{\hat{e}}_2
              \mbs{\nabla}_1\phi^{\text{eff}}_{\mb{\hat{e}}_1\mb{\hat{e}}_2}(\mb{r}_2 - \mb{r}_1)
              \psi^{(2)}_{\mb{\hat{e}}_1\mb{\hat{e}}_2}(\mb{r}_1,\mb{r}_2,t)
    \Big);
 &&
\end{flalign*} 
\noindent and for Eq. (\ref{eqn:M1Omega})
\begin{flalign*}
 \label{eqn:OneBodyM1Omega}
  \partial_t \left(\psi \mbs{\bar{\omega}} \right)
   = 
  &- \mbs{\nabla}_1 \cdot 
     \left(\int\mathrm{d}\mb{v}_1\mathrm{d}\mbs{\omega}_1
            f \mb{v}_1 \otimes \mbs{\omega}_1 
     \right)
  - \rot_1 \cdot 
     \left(\int\mathrm{d}\mb{v}_1\mathrm{d}\mbs{\omega}_1
            f \mbs{\omega}_1 \otimes \mbs{\omega}_1 
     \right)\numberthis \\
  &+ \frac{\gamma_r}{I}
     \Big( \left[-\mbs{\bar{\omega}} - 
            \gamma_r^{-1} \rot_1 V(\mb{r}_1,\mb{\hat{e}}_1) 
           \right]\psi
           -\gamma_r^{-1} 
             \int\mathrm{d}\mb{r}_2\mathrm{d}\mb{\hat{e}}_2
              \rot_1\phi^{\text{eff}}_{\mb{\hat{e}}_1\mb{\hat{e}}_2}(\mb{r}_2 - \mb{r}_1)
              \psi^{(2)}_{\mb{\hat{e}}_1\mb{\hat{e}}_2}(\mb{r}_1,\mb{r}_2,t)
    \Big).
 &&
\end{flalign*} 
\noindent To arrive at the evolution equation for the momentum density, we finally integrate over the orientations $\mb{\hat{e}}_1$. Upon defining
\begin{flalign*}
 \label{eqn:MoreOneBodyDefs}
  \rho(\mb{r}_1,t) &\equiv \int \mathrm{d}\mb{\hat{e}}_1 \psi(\mb{r}_1,\mb{\hat{e}}_1,t), \\
  \mb{m}(\mb{r}_1,t) &\equiv \int \mathrm{d}\mb{\hat{e}}_1\psi(\mb{r}_1,\mb{\hat{e}}_1,t)\mb{\hat{e}}_1,\\
  \rho(\mb{r}_1,t) \mb{\bar{\bar{v}}}(\mb{r}_1,t) 
   &\equiv
  \int\mathrm{d}\mb{\hat{e}}_1\psi(\mb{r}_1,\mb{\hat{e}}_1,t)\mb{\bar{v}}(\mb{r}_1,\mb{\hat{e}}_1,t),
 \numberthis &&
\end{flalign*}
\noindent integrating over the orientations yields for Eq. (\ref{eqn:OneBodyM0})
\begin{flalign*}
 \label{eqn:CE}
  \partial_t\rho 
   = 
  - \mbs{\nabla}_1 \cdot \left(\rho\mb{\bar{\bar{v}}}\right),
 \numberthis &&
\end{flalign*}
\noindent whereas it yields for Eq. (\ref{eqn:OneBodyM1v})
\begin{flalign*}
 \label{eqn:TowardsMB}
  m \partial_t \left(\rho \mb{\bar{\bar{v}}}\right)
   =
 &- \mbs{\nabla}_1 \cdot 
      \left( m \int\mathrm{d}\mb{\hat{e}}_1\mathrm{d}\mb{v}_1\mathrm{d}\mbs{\omega}_1
                f \mb{v}_1 \otimes \mb{v}_1             
      \right)
  -\gamma_t\rho \mb{\bar{\bar{v}}} \numberthis \\
 &-\int\mathrm{d}\mb{\hat{e}}_1
    \mbs{\nabla}_1V(\mb{r}_1,\mb{\hat{e}}_1)
    \psi(\mb{r}_1,\mb{\hat{e}}_1,t)
  +\gamma_t v_0 \mb{m}
  +\int\mathrm{d}\mb{\hat{e}}_1\mathrm{d}\mb{r}_2\mathrm{d}\mb{\hat{e}}_2 
    \mbs{\nabla}_1\phi^{\text{eff}}_{\mb{\hat{e}}_1\mb{\hat{e}}_2}(\mb{r}_2-\mb{r}_1)
    \psi^{(2)}_{\mb{\hat{e}}_1\mb{\hat{e}}_2}(\mb{r}_1,\mb{r}_2,t).
 &&
\end{flalign*}
\noindent Using 
\begin{flalign*}
 \label{eqn:NiceId}
  \int\mathrm{d}\mb{\hat{e}}_1\mathrm{d}\mb{v}_1\mathrm{d}\mbs{\omega}_1
   f \mb{v}_1 \otimes \mb{v}_1 
  =
  \int\mathrm{d}\mb{\hat{e}}_1\mathrm{d}\mb{v}_1\mathrm{d}\mbs{\omega}_1  
   f (\mb{v} - \mb{\bar{\bar{v}}})\otimes(\mb{v} - \mb{\bar{\bar{v}}})
  + \rho \mb{\bar{\bar{v}}} \otimes \mb{\bar{\bar{v}}},
 \numberthis &&
\end{flalign*}
\noindent together with Eq. (\ref{eqn:CE}), allows one to rewrite Eq. (\ref{eqn:TowardsMB}) as
\begin{flalign*}
 \label{eqn:MB}
  m \rho \frac{D\mb{\bar{\bar{v}}}}{Dt}
  =
  - \mbs{\nabla}_1 \cdot \mb{P}(\mb{r}_1,t) 
  - \gamma_t \rho\mb{\bar{\bar{v}}}
  - \int\mathrm{d}\mb{\hat{e}}_1
     \mbs{\nabla}_1V(\mb{r}_1,\mb{\hat{e}}_1)
     \psi(\mb{r}_1,\mb{\hat{e}}_1,t)
  + \gamma_t v_0\mb{m},
 \numberthis &&
\end{flalign*}
\noindent where we defined the material derivative $D/Dt \equiv \partial_t + \mb{\bar{\bar{v}}} \cdot \mbs{\nabla}_1$, and where the pressure tensor 
\begin{flalign*}
 \label{eqn:PressureTensor}
  \mb{P}(\mb{r}_1,t)
  \equiv
  &m\int\mathrm{d}\mb{\hat{e}}_1\mathrm{d}\mb{v}_1\mathrm{d}\mbs{\omega}_1
    f (\mb{v}_1 - \mb{\bar{\bar{v}}}) \otimes (\mb{v}_1 - \mb{\bar{\bar{v}}})\\
    &- \frac{1}{2} 
      \int\mathrm{d}\mb{\hat{e}}_1\mathrm{d}\mb{r}_{12}\mathrm{d}\mb{\hat{e}}_2
       \int_0^1 \mathrm{d}u
        \mb{r}_{12} \otimes \frac{\partial}{\partial\mb{r}_{12}} 
        \phi^{\text{eff}}_{\mb{\hat{e}}_1\mb{\hat{e}}_2}(\mb{r}_{12})
        \psi^{(2)}_{\mb{\hat{e}}_1\mb{\hat{e}}_2}(\mb{r}_1-u\mb{r}_{12},\mb{r}_1+(1-u)\mb{r}_{12})
 \numberthis &&
\end{flalign*}
\noindent is defined in terms of momentum flux (with respect to the mean velocity $\mb{\bar{\bar{v}}})$, and of interaction forces, here in the Kirkwood-Irving form, as is standard for the definition of the local pressure tensor (or negative of the stress tensor) \cite{MazurdeGroot,RowlinsonWidom}.\\
	\indent Having properly defined the momentum balance (\ref{eqn:MB}) for the underdamped dynamics, we are now ready to consider the overdamped limit. As explained below Eq. (\ref{eqn:Langevin}), the separation of time scales implies that the velocities $\mb{v}_i$ and $\mbs{\omega}_i$ evolve much faster than the positions $\mb{r}_i$ and orientations $\mb{\hat{e}}_i$, respectively. Motivated by this separation of time scales, we extend the approach of Enculescu and Stark \cite{ES} and make a local-equilibrium Maxwell-Boltzmann approximation\footnote{Alternatively, the separation of time scales can be exploited to explicitly solve Eq. (\ref{eqn:InertFP}) for $f^{(N)}$ by means of a multiple time scale theory, see \cite{Falasco2016}.} \cite{Archer2008} for the $N$-body distribution function
\begin{flalign*}
 \label{eqn:LESolution}
   f^{(N)}(\mb{r}^{N},\mb{\hat{e}}^N,\mb{v}^N,\mbs{\omega}^N,t)
   \simeq
   \psi^{(N)}(\mb{r}^N,\mb{\hat{e}}^N,t)
   \left(\frac{mI\beta^2}{4\pi^2}\right)^{\frac{3N}{2}}
  \prod_i
   e^{
     -\frac{\beta m}{2}\left(\mb{v}_i-\mb{\bar{v}}_i(\mb{r}^N,\mb{\hat{e}}^N,t)\right)^2
     -\frac{\beta I}{2}\left(\mbs{\omega}_i-\mbs{\bar{\omega}}_i(\mb{r}^N,\mb{\hat{e}}^N,t)\right)^2
   },\!\!\!\!\!\!\!\!\!
 \numberthis&&
\end{flalign*}
\noindent consistent with Eq. (\ref{eqn:Defs}). We defined here $\beta \equiv (k_BT)^{-1}$. \\
	\indent The overdamped version of the Fokker-Planck equation (\ref{eqn:InertFP}) is obtained by using Eq. (\ref{eqn:LESolution}) in Eq. (\ref{eqn:OneBodyM1v}) and (\ref{eqn:OneBodyM1Omega}), and combining the result with Eq. (\ref{eqn:OneBodyM0}). Upon noting that
\begin{flalign*}
 \label{eqn:Term1}
  \int\mathrm{d}\mb{v}_1\mathrm{d}\mbs{\omega}_1f \mb{v}_1 \otimes \mb{v}_1
  \overset{(\ref{eqn:LESolution})}{=}
  \frac{k_BT}{m}\psi\Id
  +N \int\mathrm{d}\mb{r}^{(2\rightarrow N)}\mathrm{d}\mb{\hat{e}}^{(2\rightarrow N)}
      \psi^{(N)}
      \mb{\bar{v}}_1\otimes\mb{\bar{v}}_1,
 \numberthis &&
\end{flalign*}
\noindent and
\begin{flalign*}
 \label{eqn:Term2}
  \int\mathrm{d}\mb{v}_1\mathrm{d}\mbs{\omega}_1f \mbs{\omega}_1 \otimes \mb{v}_1
  \overset{(\ref{eqn:LESolution})}{=}
  N \int\mathrm{d}\mb{r}^{(2\rightarrow N)}\mathrm{d}\mb{\hat{e}}^{(2\rightarrow N)}
      \psi^{(N)}
      \mbs{\bar{\omega}}_1\otimes\mb{\bar{v}}_1,
 \numberthis &&
\end{flalign*}
\noindent Eq. (\ref{eqn:OneBodyM1v}) becomes
\begin{flalign*}
 \label{eqn:OneBodyM1vOverdamped}
 m
  &\left[
   \partial_t\left(\psi\mb{\bar{v}}\right)
   \! \! + \! N\mbs{\nabla}_1\cdot
     \left(
       \int\mathrm{d}\mb{r}^{(2\rightarrow N)}\mathrm{d}\mb{\hat{e}}^{(2\rightarrow N)}
        \psi^{(N)} \mb{\bar{v}}_1 \otimes \mb{\bar{v}}_1
     \right)
    \!\!+ \!\! N\rot_1\cdot
     \left(
       \int\mathrm{d}\mb{r}^{(2\rightarrow N)}\mathrm{d}\mb{\hat{e}}^{(2\rightarrow N)}
        \psi^{(N)} \mbs{\bar{\omega}}_1 \otimes \mb{\bar{v}}_1
     \right)
  \right]\!\!\!\!\!\!\!\numberthis \\
 &=
  \left[ -\gamma_t \mb{\bar{v}}
         -\mbs{\nabla}_1 V(\mb{r}_1,\mb{\hat{e}}_1)
         +\gamma_tv_0\mb{\hat{e}}_1
         -k_BT\mbs{\nabla}_1
  \right] \psi
  -\int\mathrm{d}\mb{r}_2\mathrm{d}\mb{\hat{e}}_2
    \mbs{\nabla}_1\phi^{\text{eff}}_{\mb{\hat{e}}_1\mb{\hat{e}}_2}(\mb{r}_2-\mb{r}_1)
    \psi^{(2)}_{\mb{\hat{e}}_1\mb{\hat{e}}_2}(\mb{r}_1,\mb{r}_2,t).
 &&
\end{flalign*}
\noindent Similarly, Eq. (\ref{eqn:OneBodyM1Omega}) becomes
\begin{flalign*}
 \label{eqn:OneBodyM1OmegaOverdamped}
 I
  &\left[
   \partial_t\left(\psi\mbs{\bar{\omega}}\right)
   +N\mbs{\nabla}_1\cdot
     \left(
       \int\mathrm{d}\mb{r}^{(2\rightarrow N)}\mathrm{d}\mb{\hat{e}}^{(2\rightarrow N)}
        \psi^{(N)} \mb{\bar{v}}_1 \otimes \mbs{\bar{\omega}}_1
     \right)
   +N\rot_1\cdot
     \left(
       \int\mathrm{d}\mb{r}^{(2\rightarrow N)}\mathrm{d}\mb{\hat{e}}^{(2\rightarrow N)}
        \psi^{(N)} \mbs{\bar{\omega}}_1 \otimes \mbs{\bar{\omega}}_1
     \right)
  \right]\\
 &\overset{(\ref{eqn:LESolution})}{=}
  \left[ -\gamma_r \mbs{\bar{\omega}}
         -\rot_1 V(\mb{r}_1,\mb{\hat{e}}_1)
         -k_BT\rot_1
  \right] \psi
  -\int\mathrm{d}\mb{r}_2\mathrm{d}\mb{\hat{e}}_2
    \rot_1\phi^{\text{eff}}_{\mb{\hat{e}}_1\mb{\hat{e}}_2}(\mb{r}_2-\mb{r}_1)
    \psi^{(2)}_{\mb{\hat{e}}_1\mb{\hat{e}}_2}(\mb{r}_1,\mb{r}_2,t).
 \numberthis &&
\end{flalign*}
\noindent In the overdamped limit ($m,I\rightarrow 0$), the left-hand sides of Eq. (\ref{eqn:OneBodyM1vOverdamped}) and (\ref{eqn:OneBodyM1OmegaOverdamped}) disappear, yielding 
\begin{flalign*}
 \label{eqn:Currents}
  \psi \mb{\bar{v}} 
  &=          
  \left[-\gamma_t^{-1}\mbs{\nabla}_1 V(\mb{r}_1,\mb{\hat{e}}_1)
        +v_0\mb{\hat{e}}_1
        -\gamma_t^{-1}k_BT\mbs{\nabla}_1
  \right] \psi
  -\gamma_t^{-1}
   \int\mathrm{d}\mb{r}_2\mathrm{d}\mb{\hat{e}}_2
    \mbs{\nabla}_1\phi^{\text{eff}}_{\mb{\hat{e}}_1\mb{\hat{e}}_2}(\mb{r}_2-\mb{r}_1)
    \psi^{(2)}_{\mb{\hat{e}}_1\mb{\hat{e}}_2}(\mb{r}_1,\mb{r}_2,t),\\
 \psi \mbs{\bar{\omega}} 
  &=          
  \left[-\gamma_r^{-1}\rot_1 V(\mb{r}_1,\mb{\hat{e}}_1)
        -\gamma_r^{-1}k_BT\rot_1
  \right] \psi
  -\gamma_r^{-1}
   \int\mathrm{d}\mb{r}_2\mathrm{d}\mb{\hat{e}}_2
    \rot_1\phi^{\text{eff}}_{\mb{\hat{e}}_1\mb{\hat{e}}_2}(\mb{r}_2-\mb{r}_1)
    \psi^{(2)}_{\mb{\hat{e}}_1\mb{\hat{e}}_2}(\mb{r}_1,\mb{r}_2,t).
 \numberthis &&
\end{flalign*}
\noindent Together with Eq. (\ref{eqn:OneBodyM0}), Eq. (\ref{eqn:Currents}) forms the Smoluchowski equation for  interacting overdamped ABPs. For the non-interacting case, upon defining $\mb{j}(\mb{r}_1,\mb{\hat{e}}_1,t) \equiv \psi(\mb{r}_1,\mb{\hat{e}}_1,t)\mb{\bar{v}}(\mb{r}_1,\mb{\hat{e}}_1,t)$ and $\mb{j}_{\mb{\hat{e}}}(\mb{r}_1,\mb{\hat{e}}_1,t) \equiv \psi(\mb{r}_1,\mb{\hat{e}}_1,t)\mbs{\bar{\omega}}(\mb{r}_1,\mb{\hat{e}}_1,t)$, this Smoluchowski equation reduces to the non-interacting Smoluchowski equation (\ref{eqn:FP}) of the main text.\\
	\indent Finally, we find the overdamped version of the momentum balance (\ref{eqn:MB}). In order to find the expression for the pressure tensor (\ref{eqn:PressureTensor}) under the approximation (\ref{eqn:LESolution}), we note that
\begin{flalign*}
  m \int\mathrm{d}\mb{\hat{e}}_1\mathrm{d}\mb{v}_1\mathrm{d}\mbs{\omega}_1
     f(\mb{v}_1-\mb{\bar{\bar{v}}})\otimes(\mb{v}_1-\mb{\bar{\bar{v}}})
  = 
  \rho k_BT \Id 
  +m N 
   \int\mathrm{d}\mb{\hat{e}}_1
       \mathrm{d}\mb{r}^{(2\rightarrow N)}\mathrm{d}\mb{\hat{e}}_2^{2\rightarrow N}
    \psi^{(N)}(\mb{r}^N,\mb{\hat{e}}^N,t)
    (\mb{\bar{v}}_1-\mb{\bar{\bar{v}}})\otimes(\mb{\bar{v}}_1-\mb{\bar{\bar{v}}}),
  &&
\end{flalign*}
\noindent such that the the momentum balance (\ref{eqn:MB}) in the overdamped limit ($m,I\rightarrow 0$) reads
\begin{flalign*}
 \label{eqn:MBOverdamped}
  0
  =
  - \mbs{\nabla}_1 \cdot \mb{P}(\mb{r}_1,t) 
  - \gamma_t \rho\mb{\bar{\bar{v}}}
  - \int\mathrm{d}\mb{\hat{e}}_1
     \mbs{\nabla}_1V(\mb{r}_1,\mb{\hat{e}}_1)
     \psi(\mb{r}_1,\mb{\hat{e}}_1,t)
  + \gamma_t v_0\mb{m},
 \numberthis &&
\end{flalign*}
\noindent with the overdamped pressure tensor given as
\begin{flalign*}
 \label{eqn:PressureTensorOverdamped}
 \mb{P}(\mb{r}_1,t)
 =
 \rho(\mb{r}_1,t)k_BT\Id
 - \frac{1}{2} 
      \int\mathrm{d}\mb{\hat{e}}_1\mathrm{d}\mb{r}_{12}\mathrm{d}\mb{\hat{e}}_2
       \int_0^1 \mathrm{d}u
        \mb{r}_{12} \otimes \frac{\partial}{\partial\mb{r}_{12}} 
        \phi^{\text{eff}}_{\mb{\hat{e}}_1\mb{\hat{e}}_2}(\mb{r}_{12})
        \psi^{(2)}_{\mb{\hat{e}}_1\mb{\hat{e}}_2}(\mb{r}_1-u\mb{r}_{12},\mb{r}_1+(1-u)\mb{r}_{12}).
 \numberthis &&
\end{flalign*}
\noindent In the non-interacting case, and in the steady state of the main text, Eq. (\ref{eqn:PressureTensorOverdamped}) reduces to Eq. (\ref{eqn:FBC}) of the main text, upon defining the particle current $\mb{J}(\mb{r}) \equiv \rho(\mb{r}) \mb{\bar{\bar{v}}}(\mb{r})$, and the local pressure $P(\mb{r}) \equiv \frac{1}{3}\text{Tr}[\mb{P}(\mb{r})] = \rho(\mb{r}) k_BT$. This justifies the interpretation of Eq. (\ref{eqn:FBC}) as a force balance.
\section{2. Derivation of the solvent force balance}
\label{sec:SolventFlow}
\noindent In this section we derive that Eq. (\ref{eqn:FBS}) of the main text is the force balance governing the solvent flow on a scale where the colloids can be regarded as a continuum. To this end, we start from the hydrodynamic problem that governs the solvent flow around a single swimmer, and coarse-grain this problem to the desired larger scale. As done throughout this Electronic Supplementary Information, Latin indices $i, j, k$ shall label particles, whereas Greek indices $\alpha, \beta, \gamma$ shall refer to the Cartesian components $x,y,z$. We apply the Einstein summation convention \emph{only} to the latter Greek indices, and only in this section. Furthermore, we use the notation $A_{(\alpha\beta)} \equiv \frac{1}{2}(A_{\alpha\beta}+A_{\alpha\beta})$ to denote the symmetrization of a tensor $\mb{A}$ with respect to its Greek indices \emph{only}.\\
	\indent To describe the solvent flow around a single swimmer, we consider a model in which the swimming is generated by a nonzero slip-velocity at the surface of the single particle. This models for example biological swimmers - so-called `squirmers' - that move by the beating motion of small flagella at their body surface, or by small body deformations \cite{Blake}, but also the swimming of active colloidal particles \cite{Anderson, StarkReview}. The hydrodynamic problem is as follows. The swimmer/particle $i$, occupying a volume $V_i$ enclosed by the surface $S_i$, is assumed to have a fixed overall shape, such that it can only undergo rigid body motion, with center-of-mass velocity $\mb{v}_i$ and angular velocity $\mbs{\omega}_i$ around its center-of-mass position $\mb{r}_i$. It swims in an ambient flow $\mb{u}^{\infty}(\mb{r})$ that is assumed to solve the Stokes equation for all $\mb{r}$ in the absence of any particles. In the fluid region $V_f$, bounded by $S_i$ and by a spherical surface $S_{\infty}$ with radius $R$ that we plan to take towards $\infty$, the fluid velocity $\mb{u}^{\text{out}}(\mb{r})$ and pressure $p^{\text{out}}(\mb{r})$ satisfy
\begin{flalign*}
 \label{eqn:HydroProblem}
  \left.
   \begin{alignedat}{1}
     \mbs{\nabla} \cdot \mb{u}^{\text{out}} &=  0\\
     - \mbs{\nabla}p^{\text{out}} + \eta \mbs{\nabla}^2 \mb{u}^{\text{out}}  &= \mb{0} 
   \end{alignedat}
  \right\}
  \quad \text{with b.c.'s}\quad
  \left\{
   \begin{alignedat}{2}
    \mb{u}^{\text{out}}(\mb{r}) &= \mb{u}_i^{\text{RBM}}(\mb{r}) + \mb{u}_i^s(\mb{r}), &&\text{ for } \mb{r} \in S_i,\\
    \mb{u}^{\text{out}}(\mb{r}) &= \mb{u}^{\infty}(\mb{r}), &&\text{ for } \mb{r} \in S_{\infty},
   \end{alignedat}
  \right.
 \numberthis &&
\end{flalign*}
\noindent where $\eta$ is the dynamic solvent viscosity, where $\mb{u}_i^{\text{RBM}}(\mb{r}) \equiv \mb{v}_i + \mbs{\omega}_i \times (\mb{r} - \mb{r}_i)$ is the surface velocity of particle $i$ due to its rigid body motion, and where $\mb{u}_i^s(\mb{x})$ is the additional slip velocity satisfying $\mb{u}_i^s(\mb{x}) \cdot \mb{\hat{n}}_i = 0$, $\mb{\hat{n}}_i$ being the normal vector pointing from particle $i$ into $V_f$. The stress tensor $\mbs{\sigma}^{\text{out}}$ of the solvent is given as $\sigma^{\text{out}}_{\alpha\beta} = -p^{\text{out}} \delta_{\alpha\beta} + 2\eta\partial_{(\alpha} u^{\text{out}}_{\beta)}$; the second equation of (\ref{eqn:HydroProblem}), known as the Stokes equation, can thus also be written as $\mbs{\nabla} \cdot \mbs{\sigma}^{\text{out}} = \mb{0}$.\\
	\indent It is not Eq. (\ref{eqn:HydroProblem}) that we shall coarse-grain, but an integral representation of these differential equations. In e.g. the book of Kim and Karilla \cite{MicroHydrodynamics}, this integral representation is derived for a rigid, non-swimming particle, i.e. for $\mb{u}_i^s(\mb{r}) = \mb{0}$, a result that we extend to a finite swimming velocity $\mb{u}_i^s(\mb{r})$. It is important to realize that Eq. (\ref{eqn:HydroProblem}) is an equation for the solvent velocity $\emph{outside}$ the particle, that we have called $\mb{u}^{\text{out}}(\mb{x})$ to emphasize this. Of course, no solvent is present inside the particle, yet it shall be convenient to formally consider an `extended' solvent velocity profile defined on both $V_f$ \emph{and} $V_i$ as
\begin{flalign*}
 \label{eqn:UExtended}
  \mb{u}(\mb{r}) = 
  \left\{
   \begin{alignedat}{2}
    &\mb{u}^{\text{out}}(\mb{r}), \quad&&\text{if } \mb{r} \in V_f,\\
    &\mb{u}_i^{\text{in}}(\mb{r}), \quad&&\text{if } \mb{r} \in V_i,
   \end{alignedat}
  \right.
 \numberthis &&
\end{flalign*}
\noindent where $\mb{u}_i^{\text{in}}(\mb{r})$ is defined as the velocity field solving the Stokes equation $\emph{inside}$ particle $i$ (i.e. in $V_i$) subject to the boundary condition 
\begin{flalign*}
 \label{eqn:BCUin}
  \mb{u}_i^{\text{in}}(\mb{r}) = \mb{u}_i^{\text{RBM}}(\mb{r}) \text{ for } \mb{r} \in S_i.
 \numberthis &&
\end{flalign*}
\noindent This velocity profile inside the particle is easily solved as $\mb{u}_i^{\text{in}}(\mb{r}) = \mb{v}_i + \mbs{\omega}_i \times (\mb{r} - \mb{r}_i)$, meaning that $\mb{u}_i^{\text{in}}(\mb{r})$ has the same functional form as $\mb{u}_i^{\text{RBM}}(\mb{r})$, yet is defined on $V_i$ rather than only on $S_i$. It is therefore clear that this $\mb{u}_i^{\text{in}}(\mb{r})$ indeed satisfies the boundary condition (\ref{eqn:BCUin}); that this $\mb{u}_i^{\text{in}}(\mb{r})$ also solves the Stokes equations follows from $\mbs{\nabla} \cdot \mb{u}_i^{\text{in}}(\mb{r}) = 0$ and $\eta \partial_{(\alpha} u^{\text{in}}_{i,\beta)}=0$, which implies that the corresponding stress tensor reads $\mbs{\sigma}_i^{\text{in}}(\mb{r}) = - p^0_i \Id$ with a spatially constant pressure $p_i^0$. In order to find an integral representation for the solvent velocity profile $\mb{u}(\mb{r})$, as defined by (\ref{eqn:UExtended}), we follow the exact same procedure as done in \cite{MicroHydrodynamics}, but for a nonzero $\mb{u}_i^s(\mb{r})$. The resulting integral representation for $\mb{u}(\mb{r})$ makes use of the Green's functions that correspond to equations (\ref{eqn:HydroProblem}). These Green's functions, $\G_{\alpha\beta}$, $\mathcal{P}_{\alpha}$, and $\Sigma_{\alpha\beta\gamma}$ are defined by
\begin{flalign*}
 \label{eqn:Greens}
  \left\{
   \begin{alignedat}{1}
    &\partial_{\alpha}\G_{\alpha\beta}(\mb{r}) = 0,\\
    &8\pi\eta \partial_{\gamma} \Sigma_{\alpha\beta\gamma}(\mb{r}) = 
     - \partial_{\alpha}\mathcal{P}_{\beta}(\mb{r})
     + \eta \mbs{\nabla}^2\G_{\alpha\beta}(\mb{r})
     = -8 \pi \eta \delta_{\alpha\beta} \delta(\mb{r}).
   \end{alignedat}
  \right.
 \numberthis &&
\end{flalign*}
\noindent The integral formulation for the velocity field $\mb{u}(\mb{r})$ then reads
\begin{flalign*}
 \label{eqn:BIFfinal}
  \mb{u}(\mb{r})
  =
  \mb{u}^{\infty}(\mb{r})  
   -\frac{1}{8\pi\eta}
     \oint_{S_i}\mathrm{d}S(\mbs{\xi})(\mbs{\sigma}^{\text{out}}(\mbs{\xi})\cdot\mb{\hat{n}}_i)
      \cdot\mbs{\G}(\mb{r}-\mbs{\xi})
   -\oint_{S_i}\mathrm{d}S(\mbs{\xi}) \mb{u}_i^s(\mbs{\xi}) \cdot \mbs{\Sigma}(\mb{r}-\mbs{\xi})\cdot\mb{\hat{n}}_i.
 \numberthis &&
\end{flalign*}
\noindent For $\mb{u}_i^s(\mb{r}) = \mb{0}$, Eq. (\ref{eqn:BIFfinal}) shows that the solvent flow can be understood as resulting from a collection of force monopoles $-\mbs{\sigma}^{\text{out}}(\mbs{\xi})\cdot\mb{\hat{n}}_i$ distributed over the surface of the particle. For nonzero $\mb{u}_i^s(\mb{r})$, the effect of the last term in (\ref{eqn:BIFfinal}) is that an additional surface distribution of force $\emph{dipoles}$ comes into play. To see this, we use $\Sigma_{\alpha\beta\gamma} = (8\pi\eta)^{-1}(-\mathcal{P}_{\beta}\delta_{\alpha\gamma} + 2\eta\partial_{(\gamma}\G_{\alpha)\beta})$ to rewrite this term as
\begin{flalign*}
 \label{eqn:DipoleTerm}   
   -\oint_{S_i}\mathrm{d}S(\mbs{\xi}) u_{i\alpha}^s(\mbs{\xi})\Sigma_{\alpha\beta\gamma}(\mb{r}-\mbs{\xi})\hat{n}_{i\gamma}
   &= \frac{1}{8\pi\eta}
      \oint_{S_i}\mathrm{d}S(\mbs{\xi})\mathcal{P}_{\beta}(\mb{r}-\mbs{\xi}) 
       \mb{u}_i^s(\mbs{\xi})\cdot\mb{\hat{n}}_i
    -\frac{1}{4\pi}
      \oint_{S_i}\mathrm{d}S(\mbs{\xi})\partial_{(\gamma}\G_{\alpha)\beta}(\mb{r}-\mbs{\xi})
       u^s_{i\alpha}(\mbs{\xi})\hat{n}_{i\gamma}\\
   &= -\frac{1}{4\pi}        
       \oint_{S_i}\mathrm{d}S(\mbs{\xi}) \lim_{\epsilon \downarrow 0} \left\{
        \left[\G_{\alpha\beta}\left(\mb{r}-(\mbs{\xi}-\epsilon\mbs{\hat{\gamma}})\right)-\G_{\alpha\beta}(\mb{r}-\mbs{\xi})\right]
        \frac{u^s_{i(\alpha}(\mbs{\xi})\hat{n}_{i\gamma)}(\mbs{\xi})}{\epsilon}\right\},
 \numberthis &&
\end{flalign*}
\noindent where we used $\mb{u}_i^s(\mbs{\xi})\cdot\mb{\hat{n}}_i(\mbs{\xi})=0$, wrote out the definition of the derivative $\partial_{\gamma}$, with $\mbs{\hat{\gamma}}$ denoting the unit vector in the $\gamma$-direction, and used $A_{(\alpha\beta)}B_{\alpha\beta} = A_{(\alpha\beta)}B_{(\alpha\beta)} = A_{\alpha\beta}B_{(\alpha\beta)}$ for any tensors $\mb{A}$ and $\mb{B}$. A summation over $\gamma$ is implied in the last line of $(\ref{eqn:DipoleTerm})$, and will be in similar terms arising from this term. The combination of Eq. (\ref{eqn:BIFfinal}) and (\ref{eqn:DipoleTerm}) shows that the solvent velocity profile $\mb{u}(\mb{r})$, as defined by Eq. (\ref{eqn:UExtended}), satisfies the problem
\begin{flalign*}
 \label{eqn:HydroProblem2}
  \begin{alignedat}{2} 
   \mbs{\nabla} \cdot \mb{u}(\mb{r}) &= 0,&& \\ 
   -\partial_{\alpha}p(\mb{r})+\eta\mbs{\nabla}^2u_{\alpha}(\mb{r}) 
   &= &&\sum_i\oint_{S_i}\mathrm{d}S(\mbs{\xi}) 
                          \sigma^{\text{out}}_{\alpha\beta}(\mbs{\xi})\hat{n}_{i\beta}(\mbs{\xi})
      \delta^3(\mb{r}-\mbs{\xi})\\
    &&&+2\eta
      \sum_i\oint_{S_i}\mathrm{d}S(\mbs{\xi})\lim_{\epsilon\downarrow 0}\left\{
       \frac{u^s_{i(\alpha}(\mbs{\xi})\hat{n}_{i\gamma)}(\mbs{\xi})}{\epsilon}   
       \left[\delta^3\left(\mb{r}-(\mbs{\xi}-\epsilon\mbs{\hat{\gamma}})\right) - \delta^3(\mb{r}-\mbs{\xi})\right]
       \right\},
  \end{alignedat}
 \numberthis &&
\end{flalign*}
\noindent where we now account for many possible particles $i$ present, and where $\mb{u}(\mb{r})$ is subject to the boundary condition $\mb{u}(\mb{r}) = \mb{u}^{\infty}(\mb{r})$ for $\mb{r} \in S_{\infty}$ (with $R \rightarrow \infty$). Eq. (\ref{eqn:HydroProblem2}) indeed shows that the velocity profile $\mb{u}(\mb{r})$ can be thought of as resulting from a distribution of force monopoles, \emph{and}, for $\mb{u}^s_i(\mb{r})\neq\mb{0}$, force dipoles distributed over the surfaces of the particles.\\

\noindent It is equation (\ref{eqn:HydroProblem2}) that we shall coarse-grain. In order to do so, we define a window $w(\mb{r})$ around $\mb{r}$, that satisfies $\int\mathrm{d}\mb{r}w(\mb{r})=1$, and whose `width' determines the coarse-graining scale $L$. For definiteness, we shall take
\begin{flalign*}
 \label{eqn:Window}
  w(\mb{r}) = \frac{1}{L^3}\prod_{\alpha=x,y,z}\Theta(\frac{L}{2}-|r_{\alpha}|),
 \numberthis &&
\end{flalign*}
\noindent such that $w(\mb{r})$ is only nonzero (and equal to $L^{-3}$) inside a cube with ribbons of length $L$ centered at $\mb{r}$, that we shall refer to as $\mathcal{C}(\mb{r},L)$. We assume the window to `contain' many colloids; for our cubical window (\ref{eqn:Window}) we thus assume $L \gg a,b$, where a is the particle radius and $b$ the typical particle separation. We define the coarse-grained version of any solvent property $f(\mb{r})$ as
\begin{flalign*}
 \label{eqn:CG}
  \langle f \rangle(\mb{r})  
  = \int\limits_{V^+_f} \mathrm{d}\mb{r'} w(\mb{r}-\mb{r'}) f(\mb{r'})
  = \frac{1}{L^3}
    \int\limits_{\mathcal{C}(\mb{r},L) \cap V_f^+} \mathrm{d}\mb{r'} 
      f(\mb{r'}),
 \numberthis &&
\end{flalign*}
\noindent where $V_f^+$ denotes the fluid volume $V_f$, plus, for every particle $i$, a thin shell of width $\delta$ enclosing the particle surface $S_i$ \footnote{Formally we take $\delta \rightarrow 0$, while ensuring that $\delta > \epsilon$ at all times. All the monopoles \emph{and} dipoles appearing in Eq. (\ref{eqn:HydroProblem2}) are thus entirely contained in $V_f^+$.}. Note that the integration is \emph{not} over the volume \emph{inside} the particles, while we \emph{do} divide by the entire window volume $L^3$. Consequently, the coarse-grained solvent velocity $\langle \mb{u} \rangle (\mb{r})$ is the \emph{physical} velocity $\mb{u}^{\text{out}}(\mb{r})$ \emph{volume-averaged} over $\mathcal{C}(\mb{r},L)$. It is related to the average velocity \emph{per solvent particle} $\mb{u}^{\text{av}}(\mb{r})$ as $\langle \mb{u} \rangle (\mb{r}) = (1-\phi(\mb{r})) \mb{u}^{\text{av}}(\mb{r})$, where $\phi(\mb{r})$ is the local volume fraction of colloids.\\
	\indent We now coarse-grain Eq. (\ref{eqn:HydroProblem2}), i.e. we calculate $\langle (\ref{eqn:HydroProblem2}) \rangle (\mb{r})$. First, we note that any distribution $f(\mb{r})$ satisfies 
\begin{flalign*}
 \label{eqn:GradRule}
  \langle \mbs{\nabla}f \rangle (\mb{r})
  &= \int_{V_f^+}\mathrm{d}\mb{r'}w(\mb{r}-\mb{r'})\mbs{\nabla'}f(\mb{r'})
  =-\int_{V_f^+}\mathrm{d}\mb{r'}\mbs{\nabla'}w(\mb{r}-\mb{r'})f(\mb{r'})
  = \mbs{\nabla} \int_{V_f^+}\mathrm{d}\mb{r'}w(\mb{r}-\mb{r'})f(\mb{r'})\\
  &= \mbs{\nabla} \langle f \rangle(\mb{r}).
 \numberthis &&
\end{flalign*}
\noindent Using this property, the left-hand side of the coarse-grained version of Eq. (\ref{eqn:HydroProblem2}) becomes
\begin{flalign*}
 \label{eqn:HydroProblemCGLHS}
   -\partial_{\alpha}\langle p\rangle (\mb{r}) + \eta \mbs{\nabla}^2\langle u_{\alpha}\rangle(\mb{r}).
 \numberthis &&
\end{flalign*}
\noindent The coarse-grained version of the first term on the right-hand side of Eq. (\ref{eqn:HydroProblem2}) is
\begin{flalign*}
 \label{eqn:HydroProblemCGRHS1}
   \sum_i\oint_{S_i}\mathrm{d}S(\mbs{\xi})w(\mb{r}-\mbs{\xi})\sigma^{\text{out}}_{\alpha\beta}(\mbs{\xi})\hat{n}_{i\beta}(\mbs{\xi})
   &\approx 
     \frac{1}{L^3} \sum_{i\in\mathcal{C}(\mb{r},L)} 
      \oint_{S_i}\mathrm{d}S(\mbs{\xi})
        \sigma^{\text{out}}_{\alpha\beta}(\mbs{\xi}) \hat{n}_{i\beta}(\mbs{\xi})\\
   &=
     \frac{1}{L^3}\sum_{i\in\mathcal{C}(\mb{r},L)}F^H_{i,\alpha},
 \numberthis &&
\end{flalign*}
\noindent where $\mb{F}^H_i = \oint_{S_i}\mathrm{d}S\mbs{\sigma}^{\text{out}}\cdot\mb{\hat{n}}$ is the hydrodynamic force exerted on particle $i$, and where we neglected any contributions from particles contained only partially in $\mathcal{C}(\mb{r},L)$, which is justified by virtue of the assumption $L \gg a,b$. The coarse-grained version of the second term on the right-hand-side of Eq. (\ref{eqn:HydroProblem2}) is
\begin{flalign*}
 \label{eqn:HydroProblemCGRHS2}
  2\eta \int_{V_f^+} \mathrm{d}\mb{r'}&w(\mb{r}-\mb{r'})
   \sum_i \oint_{S_i}\mathrm{d}S(\mbs{\xi}) \lim_{\epsilon \downarrow 0}
   \left\{ \frac{u^s_{i(\alpha}(\mbs{\xi})\hat{n}_{\gamma)}(\mbs{\xi})}{\epsilon}
     \left[ \delta^3(\mb{r'}-(\mbs{\xi}-\epsilon \mbs{\hat{\gamma}}))
           -\delta^3(\mb{r'}-\mbs{\xi})
     \right]
   \right\}\\
  &=
  2\eta\sum_i \oint_{S_i} \mathrm{d}S(\mbs{\xi})
   \lim_{\epsilon\downarrow 0}
   \left\{ 
     \frac{w(\mb{r}-(\mbs{\xi}-\epsilon \mbs{\hat{\gamma}}))-w(\mb{r}-\mbs{\xi})}{\epsilon}
   \right\}
   u^s_{i(\alpha}(\mbs{\xi})\hat{n}_{i\gamma)}(\mbs{\xi})\\
  &= 2\eta \sum_i\oint_{S_i} \mathrm{d}S(\mbs{\xi})
      \frac{\partial w(\mb{r}-\mbs{\xi}) }{\partial r_{\gamma}}
        u^s_{i(\alpha}(\mbs{\xi})\hat{n}_{i\gamma)}(\mbs{\xi})\\
  &\overset{\mathclap{(\ref{eqn:Window})}}{= }
    -\frac{2\eta}{L^3} \sum_i \oint_{S_i} \mathrm{d}S(\mbs{\xi})
      \left(\prod_{\beta \neq \gamma} \Theta(\frac{L}{2} - |r_{\beta} - \xi_{\beta}|)\right)
      \left\{ \delta\big(r_{\gamma} - (\xi_{\gamma} + \frac{L}{2})\big)
            -\delta\big(r_{\gamma}-(\xi_{\gamma}-\frac{L}{2})\big)
      \right\}   
     u^s_{i(\alpha}(\mbs{\xi})\hat{n}_{i\gamma)}(\mbs{\xi}) \\
  &=-\frac{2\eta}{L^3}\sum_i
     \left(\oint_{\delta \mathcal{C}^+_{\gamma}(\mb{r},L) \cap S_i} \mathrm{d}l(\mbs{\xi})
            -\oint_{\delta \mathcal{C}^-_{\gamma}(\mb{r},L)\cap S_i}\mathrm{d}l(\mbs{\xi}) 
     \right)
      u^s_{i(\alpha}(\mbs{\xi})\hat{n}_{i\gamma)}(\mbs{\xi}),
 \numberthis &&
\end{flalign*}
\noindent where in the last line $\delta \mathcal{C}^{\pm}_{\gamma}(\mb{r},L)$ denotes the face of the $\mathcal{C}(\mb{r},L)$ cube with outward normal $\pm \mbs{\hat{\gamma}}$. The integrations in the last line thus run over the intersection of the $\delta \mathcal{C}^{\pm}_{\gamma}(\mb{r},L)$-face with the surface $S_i$ of any particle $i$ that intersects it. For any particle $i$, the integration domain is thus an intersection between two surfaces, which forms a line. The magnitude of the contributions (\ref{eqn:HydroProblemCGRHS1}) and (\ref{eqn:HydroProblemCGRHS2}) can now be estimated. Denoting the magnitude of $F_i^H$ by $F$, and the colloid density by $\rho$, such that the number of colloids inside $\mathcal{C}(\mb{r},L)$ approximately equals $\rho L^3$, the magnitude of the contribution (\ref{eqn:HydroProblemCGRHS1}) is estimated as $L^{-3} (\rho L^3) F = \rho F \approx F \phi a^{-3}$, where $\phi$ denotes the packing fraction of the colloids, and where $a$ denotes the particle size. To estimate the contribution of either integral in (\ref{eqn:HydroProblemCGRHS2}), we note that i) the number of particles intersecting $\delta \mathcal{C}^{\pm}_{\gamma}(\mb{r},L)$ has as approximate upper bound $(\rho L^3)^{\frac{2}{3}}$ (in fact, the number of intersecting particles is much less for a dilute suspension), ii) for any particle intersecting $\delta \mathcal{C}^{\pm}_{\gamma}(\mb{r},L)$, the length of the intersection line is of the order $a$, and iii) $\eta u^s_i \approx a \sigma^{\text{out}} \approx F/a$. Therefore, the contribution of either integral in (\ref{eqn:HydroProblemCGRHS2}) is approximated as $L^{-3} (\rho L^3)^{\frac{2}{3}} a (F/a) = F \rho^{2/3}L^{-1} \approx F \phi^{\frac{2}{3}}a^{-2}L^{-1}$. As $L^{-1} \ll a^{-1}$, the contribution of $(\ref{eqn:HydroProblemCGRHS2})$ is negligible as compared to the contribution of ($\ref{eqn:HydroProblemCGRHS1})$. Therefore, the coarse-grained version of Eq. (\ref{eqn:HydroProblem2}) reads
\begin{flalign*}
 \label{eqn:HydroProblemCG}
  -\partial_{\alpha} \langle p \rangle (\mb{r}) + \eta \mbs{\nabla}^2 \langle u_{\alpha} \rangle (\mb{r})
  =
  \frac{1}{L^3} \sum\limits_{i \in \mathcal{C}(\mb{r},L)} F^H_{i,\alpha}.
 \numberthis&&
\end{flalign*}
	\indent The hydrodynamic force $\mb{F}^H_i$ experienced by a spherical particle $i$ can be decomposed as $\mb{F}_i^H = - \gamma_t\mb{v}_i + \gamma_t v_0 \mb{\hat{e}}_i$, where $\mb{v}_i$ is the velocity of particle $i$ and $\mb{\hat{e}}_i$ its orientation \cite{BradyBodyForce}. The evolution of $\mb{v}_i(t)$ and $\mb{\hat{e}}_i(t)$ are governed by the Langevin dynamics of the particles \footnote{This is under the assumption that the effect of the slip velocity $\mb{u}_i^s$ is to displace particle $i$ only translationally; if it also rotates the particle an additional `self-torque' has to be added to the Langevin equations.}. Consequently, even though it was left implicit so far, the solvent pressure $\langle p \rangle (\mb{r},t)$ and velocity $\langle \mb{u} \rangle (\mb{r},t)$ actually depend on time, via Eq. (\ref{eqn:HydroProblemCG}). To relate the right-hand side of Eq. (\ref{eqn:HydroProblemCG}) to the probability distribution function $\psi(\mb{r},\mb{e},t)$ of the particles - whose time evolution is governed by the Smoluchowski equation - we assume the dynamics of the particles not to change significantly throughout a window, meaning that e.g. the external potential $V^{\text{ext}}(\mb{r})$ must not vary significantly under $r_{\alpha} \rightarrow r_{\alpha} + L$ \footnote{In the main text we \emph{do} consider a membrane potential that changes on the scale $a \ll L$. However, this is in a planar geometry; if one employs a window that is thin in the direction perpendicular to the membrane, and elongated in the parallel direction(s), it can still contain many colloids, yet have an approximately constant $V(\mb{r})$ inside.}. This implies that $\psi$ is approximately constant within any window, i.e. $\psi(\mb{r'},\mb{e},t) \approx \psi(\mb{r},\mb{e},t)$ for $\mb{r'} \in \mathcal{C}(\mb{r},L)$, for any $\mb{r}$. In this case, the sum over all particles in $\mathcal{C}(\mb{r},L)$ (which are many) coincides with a sum over different realizations of the noise appearing in the Langevin equation, such that
the coarse-grained Stokes equation (\ref{eqn:HydroProblemCG}) becomes
\begin{flalign*}
 \label{eqn:CG1}
  -\mbs{\nabla}\langle p\rangle (\mb{r},t) + \eta \mbs{\nabla}^2 \langle \mb{u}\rangle(\mb{r},t)
  = & \int\mathrm{d}\mb{\hat{e}} \psi(\mb{r},\mb{\hat{e}},t)
       \left[ - \gamma_t\mb{\bar{v}}(\mb{r},\mb{\hat{e}},t)
              + \gamma_tv_0\mb{\hat{e}}
       \right]\\
  \overset{(\ref{eqn:MoreOneBodyDefs})}{=}  
     \! \! \! \!& \ \ - \gamma_t \rho(\mb{r},t) \mb{\bar{\bar{v}}}(\mb{r},t) 
   + \gamma_t v_0 \mb{m}(\mb{r},t)\\
  \equiv & \ \mb{f}^f(\mb{r},t) + \mb{f}^p(\mb{r},t),
 \numberthis &&
\end{flalign*}
\noindent where we defined the frictional body force $\mb{f}^f(\mb{r},t)\equiv-\gamma_t\rho(\mb{r},t)\mb{\bar{\bar{v}}}(\mb{r},t)$ and the propulsion body force $\mb{f}^p(\mb{r},t)\equiv\gamma_tv_0\mb{m}(\mb{r},t)$, which are the internal body forces appearing in the colloidal force balance (\ref{eqn:FBC}) of the main text. Note that in the main text we simply denoted $\langle \mb{u}(\mb{r},t) \rangle$ by $\mb{u}(\mb{r},t)$, and $\langle p(\mb{r},t) \rangle$ by $P_s(\mb{r},t)$.\\
	\indent Both Eq. (\ref{eqn:HydroProblemCG}) and (\ref{eqn:CG1}) show (upon bringing all the terms to the left-hand side) that on the coarse-grained scale the solvent flow is simply governed by the Stokes equation, equipped with body forces equal to the opposite of the hydrodynamic forces experienced by the particles. We remark here that in the overdamped limit, the Langevin equation (\ref{eqn:Langevin}) reduces to $\mb{0} = \mb{F}^H_i-\mbs{\nabla}_i(V+U^{\text{eff}})+\sqrt{2\gamma_tk_BT}\mbs{\eta}^t_i$, showing that the hydrodynamic, external, interaction and Brownian forces acting on any particle exactly balance. It is therefore only when $-\mbs{\nabla}_i(V+U^{\text{eff}})+\sqrt{2\gamma_tk_BT}\mbs{\eta}^t_i=\mb{0}$, that the hydrodynamic force $\mb{F}^H_i=\mb{0}$, and that the motion is usually referred to as `force-free' \cite{BradyBodyForce}. However, when $-\mbs{\nabla}_i(V+U^{\text{eff}})+\sqrt{2\gamma_tk_BT}\mbs{\eta}^t_i\neq\mb{0}$, the hydrodynamic force $\mb{F}^H_i\neq\mb{0}$ contributes to Eq. (\ref{eqn:HydroProblemCG}). As the stochastic force $\mbs{\eta}_i^t$ time-averages to zero, the essential factor distinguishing these two cases is whether $-\mbs{\nabla}_i(V+U^{\text{eff}})$ is nonzero. In the setting of the main text, where $U^{\text{eff}}$ was neglected for the dilute suspension, the fact that the hydrodynamic force is nonzero (i.e. the fact that $\mb{f}^f + \mb{f}^p \neq \mb{0}$) near the membrane, is in this sense a consequence of the external force $-\mbs{\nabla}_iV$ exerted on the colloids by the membrane.
\section{3. Osmotic pressure with interactions}
\label{sec:IntOsmo}
\noindent This section shows the conclusion of the main paper - that the activity-induced increase in osmotic pressure can be attributed to an increase in the chemical potential of the solvent - to hold true also in the presence of interactions.\\
	\indent We start by writing the force balance of the overdamped colloids (\ref{eqn:MBOverdamped}) as
\begin{flalign*}
 \label{eqn:IntFBC}
  \mb{0}
  =
  \mb{f}^e(\mb{r},t) + \mb{f}^f(\mb{r},t) + \mb{f}^p(\mb{r},t) 
  - \mbs{\nabla} \cdot \mb{P}(\mb{r},t),
 \numberthis &&
\end{flalign*}
\noindent where we defined the \emph{external} body force $\mb{f}^e(\mb{r},t) \equiv -\int\mathrm{d}\mb{\hat{e}}\mbs{\nabla}V(\mb{r},\mb{\hat{e}})\psi(\mb{r},\mb{\hat{e}},t)$, and where we recall the definitions of the frictional body force $\mb{f}^f(\mb{r},t)=-\gamma_t\rho(\mb{r},t)\mb{\bar{\bar{v}}}(\mb{r},t)$ and the propulsion body force $\mb{f}^p(\mb{r},t)=\gamma_tv_0\mb{m}(\mb{r},t)$, both of which are \emph{internal}. Furthermore, we recall from section 1 that
\begin{flalign*}
 \label{eqn:PRelation}
  - \mbs{\nabla}_1\cdot\mb{P}(\mb{r}_1,t)
  =
  - \mbs{\nabla}_1 \rho(\mb{r}_1,t)k_BT
  - \int\mathrm{d}\mb{\hat{e}}_1\mathrm{d}\mb{r}_2\mathrm{d}\mb{\hat{e}}_2
     \mbs{\nabla}_1 \phi^{\text{eff}}_{\mb{\hat{e}}_1\mb{\hat{e}}_2}(\mb{r}_2-\mb{r}_1)
     \psi^{(2)}_{\mb{\hat{e}}_1\mb{\hat{e}}_2}(\mb{r}_1,\mb{r}_2,t).
 \numberthis &&
\end{flalign*}
\noindent The second term in (\ref{eqn:PRelation}) represents an effective force exerted on colloids at $\mb{r}_1$ by other colloids. These effective interactions are due to both direct colloid-colloid interactions and solvent induced interactions, which can be made explicit as
\begin{flalign*}
 \label{eqn:RewriteEffectiveInteractions}  
  - \int\mathrm{d}\mb{\hat{e}}_1 \mathrm{d}\mb{r}_2\mathrm{d}\mb{\hat{e}}_2  
     \mbs{\nabla}_1\phi^{\text{eff}}_{\mb{\hat{e}}_1\mb{\hat{e}}_2}(\mb{r_2}-\mb{r}_1)
     \psi^{(2)}_{\mb{\hat{e}_1}\mb{\hat{e}}_2}(\mb{r}_1,\mb{r}_2,t)
  =
  &- \int\mathrm{d}\mb{\hat{e}}_1 \mathrm{d}\mb{r}_2\mathrm{d}\mb{\hat{e}}_2  
     \mbs{\nabla}_1\phi_{\mb{\hat{e}}_1\mb{\hat{e}}_2}(\mb{r}_2-\mb{r}_1)
     \psi^{(2)}_{\mb{\hat{e}}_1\mb{\hat{e}}_2}(\mb{r}_1,\mb{r}_2,t)\\
  &- \int\mathrm{d}\mb{\hat{e}}_1 \mathrm{d}\mb{r}_2
     \mbs{\nabla}_1\phi_{\mb{\hat{e}}_1s}(\mb{r}_2-\mb{r}_1)
     \psi^{(2)}_{\mb{\hat{e}}_1s}(\mb{r}_1,\mb{r}_2,t).
\numberthis&&
\end{flalign*}
\noindent Here $\phi_{\mb{\hat{e}}_1\mb{\hat{e}}_2}(\mb{r}_2-\mb{r}_1)$ and $\phi_{\mb{\hat{e}}_1s}(\mb{r}_2-\mb{r}_1)$ denote the bare colloid-colloid and the bare colloid-solvent interaction pair potential, respectively. Furthermore, $\psi^{(2)}_{\mb{\hat{e}}_1s}(\mb{r}_1,\mb{r}_2,t)$ denotes the two-body colloid-solvent distribution function.\\
	\indent Next, we consider the momentum balance for the solvent, i.e. the analogue of Eq. (\ref{eqn:MB}) for the solvent. As we consider a regime in which the Reynolds number of the solvent $Re \ll 1$, the inertial terms in the solvent momentum balance are negligible. Therefore, the various body forces acting on the solvent, to wit, a possible external body force $\mb{f}^e_s(\mb{r},t)$, \emph{and} the negative of the internal forces $\mb{f}^f(\mb{r},t)$ and $\mb{f}^p(\mb{r},t)$ that act on the colloids, have to balance the divergence of the solvent momentum flux tensor $\mbs{\mathcal{J}}^{\text{mom}}_s(\mb{r},t)$, and the solvent-colloid and solvent-solvent interaction forces:
\begin{flalign*}
 \label{eqn:IntFBS}
  \mb{0}
  = \  
  & \mb{f}^e_s(\mb{r}_1,t)
  - \mb{f}^f(\mb{r}_1,t)
  - \mb{f}^p(\mb{r}_1,t)\numberthis \\
  &- \mbs{\nabla}_1 \cdot \mbs{\mathcal{J}}^{\text{mom}}_s(\mb{r}_1,t)
  - \int\mathrm{d}\mb{r}_2\mathrm{d}\mb{\hat{e}}_2 
     \mbs{\nabla}_1\phi_{s\mb{\hat{e}}_2}(\mb{r}_2-\mb{r}_1)
     \psi^{(2)}_{s\mb{\hat{e}}_2}(\mb{r}_1,\mb{r}_2)
  - \int\mathrm{d}\mb{r}_2
     \mbs{\nabla}_1\phi_{ss}(\mb{r}_2-\mb{r}_1)
     \psi^{(2)}_{ss}(\mb{r}_1,\mb{r}_2).
 &&
\end{flalign*}
\noindent The momentum flux tensor $\mbs{\mathcal{J}}^{\text{mom}}_s(\mb{r}_1,t)$ of the solvent is defined, analogous to the colloidal momentum flux (first term on the right-hand side of Eq. (\ref{eqn:PressureTensor})), as
\begin{flalign*}
 \label{eqn:SMomFlux}
   \mbs{\mathcal{J}}^{\text{mom}}_s(\mb{r}_1,t) 
   =
   m \int\mathrm{d}\mb{v}_1 f_s(\mb{r}_1,\mb{v}_1,t)(\mb{v}_1-\mb{u})\otimes(\mb{v}_1-\mb{u}).
 \numberthis &&
\end{flalign*}
\noindent Here the probability density $f_s(\mb{r}_1,\mb{v}_1,t)$ to find a solvent particle at position $\mb{r}_1$ with velocity $\mb{v}_1$ at time $t$ defines the solvent number density $\mb{\rho}_s(\mb{r}_1,t)$ and mean solvent velocity $\mb{u}(\mb{r}_1,t)$ as
\begin{flalign*}
 \label{eqn:SolventDefs}
  \rho_s(\mb{r}_1,t)
  &\equiv
  \int\mathrm{d}\mb{v}_1 f_s(\mb{r}_1,\mb{v}_1,t),\\
  \rho_s(\mb{r}_1,t)\mb{u}(\mb{r}_1,t)
  &\equiv
  \int\mathrm{d}\mb{v}_1 f_s(\mb{r}_1,\mb{v}_1,t)\mb{v}_1.
 \numberthis &&
\end{flalign*}
\noindent In equilibrium (where $\mb{u} = \mb{0}$), the equipartition theorem implies $\mbs{\mathcal{J}}_s^{\text{mom}}(\mb{r},t) = \rho_s(\mb{r})k_BT$. The divergence of this momentum flux then combines with the interaction terms in Eq. (\ref{eqn:IntFBS}), characterized by the equilibrium two-body distribution functions $\psi^{(2)}_{s\mb{\hat{e}}_2}(\mb{r}_1,\mb{r}_2) = \psi^{(2),\text{eq}}_{s\mb{\hat{e}}_2}(\mb{r}_1,\mb{r}_2)$ and $\psi^{(2)}_{ss}(\mb{r}_1,\mb{r}_2) = \psi^{(2),\text{eq}}_{ss}(\mb{r}_1,\mb{r}_2)$, to give \cite{DFT}
\begin{flalign*}
 \label{eqn:DFTRelation}
  -\mbs{\nabla}_1\rho_s(\mb{r}_1)&k_BT
  - \int\mathrm{d}\mb{r}_2\mathrm{d}\mb{\hat{e}}_2 
     \mbs{\nabla}_1\phi_{s\mb{\hat{e}}_2}(\mb{r}_2-\mb{r}_1)
     \psi^{(2),\text{eq}}_{s\mb{\hat{e}}_2}(\mb{r}_1,\mb{r}_2)
  - \int\mathrm{d}\mb{r}_2
     \mbs{\nabla}_1\phi_{ss}(\mb{r}_2-\mb{r}_1)
     \psi^{(2),\text{eq}}_{ss}(\mb{r}_1,\mb{r}_2)\\
  &=
  -\rho_s(\mb{r}_1)\mbs{\nabla}_1\mu_s^{\text{int}}(\mb{r}_1),
 \numberthis &&
\end{flalign*}
\noindent where the \emph{intrinsic chemical potential of the solvent} is defined as $\mu_s^{\text{int}}(\mb{r}_1) \equiv \delta F[\psi,\rho_s]/\delta\rho_s(\mb{r}_1)$, $F[\psi,\rho_s]$ being the free energy functional of the colloid-solvent mixture. To proceed out of equilibrium we follow Archer \cite{Archer2008}. Out of equilibrium, the momentum flux tensor generally receives an additional contribution, whose divergence can, under suitable approximations, be written as $-\eta^{(K)}\mbs{\nabla}^2\mb{u}$, such that
\begin{flalign*}
 \label{eqn:SMomFlux2}
 -\mbs{\nabla} \cdot \mbs{\mathcal{J}}^{\text{mom}}_s(\mb{r},t) 
   =
 -\mbs{\nabla}\rho_s(\mb{r},t)k_BT 
 + \eta^{(K)}\mbs{\nabla}^2\mb{u}(\mb{r},t).
 \numberthis &&
\end{flalign*}
\noindent Also the interaction terms in Eq. (\ref{eqn:IntFBS}) give additional contributions as compared to equilibrium, due to deviations of the correlation functions from their equilibrium values. These extra contributions can, under similar approximations, be shown to be $\eta^{(V)}\mbs{\nabla}^2\mb{u}$. Together, these two out-of-equilibrium corrections add up to $\eta\mbs{\nabla}^2\mb{u}$, where the viscosity $\eta = \eta^{(K)}+\eta^{(V)}$ comprises both kinetic contributions and contributions from interactions, respectively. For details see Archer \cite{Archer2008} and Kreuzer \cite{Kreuzer}. Putting it all together, the solvent force balance (\ref{eqn:IntFBS}) thus reads
\begin{flalign*}
 \label{eqn:IntFBS2}
  \mb{0}
  =   
   \mb{f}^e_s(\mb{r}_1,t)
  - \mb{f}^f(\mb{r}_1,t)
  - \mb{f}^p(\mb{r}_1,t)
  - \rho_s(\mb{r}_1,t) \mbs{\nabla}_1 \mu_s^{\text{int}}(\mb{r}_1,t) + \eta \mbs{\nabla}_1^2\mb{u}(\mb{r}_1,t).
   \numberthis &&
\end{flalign*}
\noindent As an aside, for the case of no colloid-solvent interactions, i.e. as in the main text (where we neglected any effective colloid-colloid interactions, and hence, according to Eq. (\ref{eqn:RewriteEffectiveInteractions}), also colloid-solvent interactions), the term $-\rho_s \mbs{\nabla}_1\mu_s^{\text{int}}$ can be written as 
\begin{flalign*}
 \label{eqn:Simplification}
  -\rho_s(\mb{r}_1,t) \mbs{\nabla}_1 \mu_s^{\text{int}}(\mb{r}_1,t)
  \overset{\phi_{s\mb{\hat{e}}}=0}{=}
  -\mbs{\nabla}_1\rho_s(\mb{r}_1,t) k_BT
  -\int\mathrm{d}\mb{r}_2
    \mbs{\nabla}_1\phi_{ss}(\mb{r}_2-\mb{r}_1)
    \psi^{(2),\text{eq}}_{ss}(\mb{r}_1,\mb{r}_2)
  =
  \mathrlap{-\mbs{\nabla} \cdot \mbs{P}_s(\mb{r}_1,t),}
 \numberthis &&
\end{flalign*}
i.e. as minus the divergence of the solvent pressure tensor
\footnote{
There is some ambiguity in what quantity to call the solvent pressure tensor here. We could also define the solvent pressure tensor
\begin{equation*}
 \quad\tilde{\mb{P}}_s(\mb{r},t) 
  \equiv 
 \mbs{\mathcal{J}}_s^{\text{mom}}(\mb{r},t) 
 -\frac{1}{2}
   \int\mathrm{d}\mb{r}_{12}\int_0^1\mathrm{d}u
    \mb{r}_{12}\otimes\frac{d}{d\mb{r}_{12}}\phi_{ss}(\mb{r}_{12})
    \psi^{(2)}_{ss}(\mb{r}_1-u\mb{r}_{12},\mb{r}_1+(1-u)\mb{r}_{12}),
\end{equation*}
\noindent which is more analogous to the pressure tensor of the total suspension $\mb{P}_{\text{tot}}(\mb{r},t)$ defined in Eq. (\ref{eqn:PtotI}), and in terms of which the solvent force balance without colloid-solvent interactions would read
\begin{equation*}
 \quad\mb{0} = \mb{f}^e_s(\mb{r},t)-\mb{f}^p(\mb{r},t)-\mbs{\nabla}\cdot\tilde{\mb{P}}_s(\mb{r},t).
\end{equation*}
\noindent The two solvent pressure tensors are related as $-\mbs{\nabla}\cdot\tilde{\mbs{P}}_s(\mb{r},t)=-\mbs{\nabla}\cdot\mbs{P}_s(\mb{r},t)+\eta\nabla^2\mb{u}(\mb{r},t)$.
}
\begin{flalign*}
 \label{eqn:SPressureTensor}
  \mb{P}_s(\mb{r}_1,t)
  =
  \rho_s(\mb{r}_1,t) k_BT \Id
  - \frac{1}{2} 
    \int\mathrm{d}\mb{r}_{12}\int_0^1\mathrm{d}u
     \mb{r}_{12} \otimes \frac{\partial}{\partial \mb{r}_{12}} 
     \phi_{ss}(\mb{r}_{12}) 
     \psi^{(2),\text{eq}}_{ss}(\mb{r}_1-u\mb{r}_{12},\mb{r}_1+(1-u)\mb{r}_{12}),
 \numberthis &&
\end{flalign*}
\noindent that contains the solvent-solvent interactions in Kirkwood-Irving form \cite{RowlinsonWidom}. Upon assuming the pressure tensor (\ref{eqn:SPressureTensor}) to be isotropic, the solvent force balance (\ref{eqn:IntFBS2}) then reduces to the solvent force balance (\ref{eqn:FBS}) of the main text (and Eq. (\ref{eqn:CG1}) of this Electronic Supplementary Information). In the presence of arbitrary solvent-colloid interactions however, an identification akin to (\ref{eqn:SPressureTensor}) is not generally possible, and we work with Eq. (\ref{eqn:IntFBS2}) as the solvent force balance.\\
	\indent The force balance of the complete suspension is obtained by adding the colloid force balance (\ref{eqn:IntFBC}) (using Eq. (\ref{eqn:PRelation}) and (\ref{eqn:RewriteEffectiveInteractions})) and the solvent force balance in the form of (\ref{eqn:IntFBS}), as
\begin{flalign*}
  \begin{alignedat}{2}
  0 &= \mb{f}^e(\mb{r}_1,t)
     +\mb{f}^e_s(\mb{r}_1,t)
     - \mbs{\nabla}_1\cdot\Big(\rho(\mb{r}_1,t)k_BT\Id
                               +\mbs{\mathcal{J}}_s^{\text{mom}}(\mb{r}_1,t)
                          \Big)
     &&- \int\mathrm{d}\mb{\hat{e}}_1\mathrm{d}\mb{r}_2\mathrm{d}\mb{\hat{e}}_2
        \mbs{\nabla}_1\phi_{\mb{\hat{e}}_1\mb{\hat{e}}_2}(\mb{r}_2-\mb{r}_1)
        \psi^{(2)}_{\mb{\hat{e}}_1\mb{\hat{e}}_2}(\mb{r}_1,\mb{r}_2)\\
     &&&- \int\mathrm{d}\mb{\hat{e}}\mathrm{d}\mb{r}_2
        \left( \mbs{\nabla}_1\phi_{\mb{\hat{e}}s}(\mb{r}_2-\mb{r}_1)
               \psi^{(2)}_{\mb{\hat{e}}s}(\mb{r}_1,\mb{r}_2)
               + \mb{\hat{e}} \leftrightarrow s
        \right)\\
     &&&- \int \mathrm{d}\mb{r}_2
         \mbs{\nabla}_1\phi_{ss}(\mb{r}_2-\mb{r}_1)
         \psi^{(2)}_{ss}(\mb{r}_1,\mb{r}_2) \\
   &= 
    \mathrlap
    {
     \mb{f}^e(\mb{r}_1,t)
     +\mb{f}^e_s(\mb{r}_1,t)     
     - \mbs{\nabla}_1\cdot \mb{P}_{\text{tot}}(\mb{r}_1,t),
    }
  \end{alignedat}
  \label{eqn:IntFBTot}
  \numberthis&&
\end{flalign*}
\noindent where the pressure tensor of the total suspension $\mb{P}_{\text{tot}}(\mb{r}_1,t)$ includes the momentum flux of the colloids and of the solvent, and all colloid-colloid, colloid-solvent and solvent-solvent interactions in Kirkwood-Irving form \cite{RowlinsonWidom}, as
\begin{flalign*}
 \label{eqn:PtotI}
 \begin{alignedat}{1}
   \mb{P}_{\text{tot}}(\mb{r}_1,t)
   =
   &\Big(\rho(\mb{r}_1,t)k_BT\Id+\mbs{\mathcal{J}}_s^{\text{mom}}(\mb{r}_1,t)\Big)\Id\\
   &-\frac{1}{2} 
    \int \mathrm{d}\mb{\hat{e}}_1\mathrm{d}\mb{r}_{12}\mathrm{d}\mb{\hat{e}_2}\int_0^1\mathrm{d}u
    \mb{r}_{12}\otimes\frac{\partial}{\partial \mb{r}_{12}}
    \phi_{\mb{\hat{e}}_1\mb{\hat{e}}_2}(\mb{r}_{12})
    \psi^{(2)}_{\mb{\hat{e}}_1\mb{\hat{e}}_2}(\mb{r}_1-u\mb{r}_{12},\mb{r}_1+(1-u)\mb{r}_{12})\\
   &-\frac{1}{2} 
    \int \mathrm{d}\mb{\hat{e}}\mathrm{d}\mb{r}_{12}\int_0^1\mathrm{d}u
    \Big(
      \mb{r}_{12}\otimes\frac{\partial}{\partial \mb{r}_{12}}
      \phi_{\mb{\hat{e}}s}(\mb{r}_{12})
      \psi^{(2)}_{\mb{\hat{e}}s}(\mb{r}_1-u\mb{r}_{12},\mb{r}_1+(1-u)\mb{r}_{12})
      +\mb{\hat{e}} \leftrightarrow s 
    \Big)\\
   &-\frac{1}{2}
    \int\mathrm{d}\mb{r}_{12}\int_0^1\mathrm{d}u
    \mb{r}_{12}\otimes\frac{\partial}{\partial \mb{r}_{12}} 
    \phi_{ss}(\mb{r}_{12})
    \psi^{(2)}_{ss}(\mb{r}_1-u\mb{r}_{12},\mb{r}_1+(1-u)\mb{r}_{12}).
 \end{alignedat}
 \numberthis&&
\end{flalign*}\\
	\indent As in the main text, we now consider a flux-free steady state, where the solvent velocity $\mb{u}(\mb{r},t)=\mb{0}$ (on the scale coarse-grained over the colloids). Adding the colloid and solvent force balance in the form of Eq. (\ref{eqn:IntFBC}) and (\ref{eqn:IntFBS2}), respectively, and comparing with the total force balance (\ref{eqn:IntFBTot}) then reveals
\begin{flalign*}
 \label{eqn:GDI}
  \mbs{\nabla} \cdot \mb{P}_{\text{tot}}(\mb{r})
  =
  \mbs{\nabla} \cdot \mb{P}(\mb{r})
  + \rho_s(\mb{r}) \mbs{\nabla} \mu_s^{\text{int}}(\mb{r}).
\numberthis&&
\end{flalign*}
	\indent In an isotropic bulk, the pressure tensor of the effective colloids-only system reduces to a scalar as $\mb{P} = P \Id$. Similarly, the total pressure tensor $\mb{P}_{\text{tot}} = P_{\text{tot}} \Id$. Furthermore, in a bulk characterized by colloid density $\rho$, solvent density $\rho_s$, and colloid propulsion speed $v_0$, all the two-body correlation functions are uniquely characterized by ($\rho$, $\rho_s$, $v_0$) (at fixed temperature). Therefore, Eq. (\ref{eqn:PressureTensorOverdamped}) and (\ref{eqn:PtotI}) reveal the pressures $P(\rho,\rho_s,v_0)$ and $P_{\text{tot}}(\rho,\rho_s,v_0)$ to be state functions in the bulk. Also the solvent chemical potential $\mu_s^{\text{int}}$ is in bulk a function of the colloid density $\rho$ and solvent density $\rho_s$, owing to its definition in terms of the free energy functional. This state function $\mu_s^{\text{int}}(\rho,\rho_s)$ can be inverted to $\rho_s(\mu_s^{\text{int}},\rho)$, such that the bulk solvent can be characterized by $\mu_s^{\text{int}}$ instead of $\rho_s$. Consequently, the bulk pressures can be expressed as $P(\rho,\mu_s^{\text{int}},v_0)$ and $P_{\text{tot}}(\rho,\mu_s^{\text{int}},v_0)$.\\
	\indent To calculate the osmotic pressure, we again specialize to the planar geometry of Fig. \ref{fig:NicePicture} of the main text, where $\mb{f}^e(\mb{r})=f^e_z(z)\mb{\hat{z}}$, and where we now also allow for a non-perfect membrane that can exert a force $\mb{f}^e_s(\mb{r})=f^e_{sz}(z)\mb{\hat{z}}$ on the solvent. The osmotic pressure then follows as 
\begin{flalign*}
\label{eqn:PwallI}
  \Pi 
  = 
  \int_{z_{\text{res}}}^{z_b} \mathrm{d}z \left( f^e_z(z) + f_{sz}^e(z) \right)
  \overset{(\ref{eqn:IntFBTot})}{=} 
  P_{\text{tot}}(\rho,\mu_s^b,v_0) - P_{\text{tot}}(\rho=0,\mu_s^{\text{res}},v_0),
\numberthis&&
\end{flalign*}
\noindent i.e. as the difference in total pressure on the opposing sides of the membrane. This difference can be obtained by integrating Eq. (\ref{eqn:GDI}) from the reservoir to the bulk at the opposing side of the membrane, to be
\begin{flalign*}
 \label{eqn:TowardsPiI}
   P_{\text{tot}}(\rho,\mu_s^b,v_0) - P_{\text{tot}}(\rho=0,\mu_s^{\text{res}},v_0)
   =
   P(\rho,\mu_s^b,v_0)
   + \int_{z_{\text{res}}}^{z_b} \mathrm{d}z \rho_s(z) \partial_z \mu_s^{\text{int}}(z).
\numberthis&&
\end{flalign*}
\begin{figure}[t]
 \includegraphics[width=0.5\linewidth]{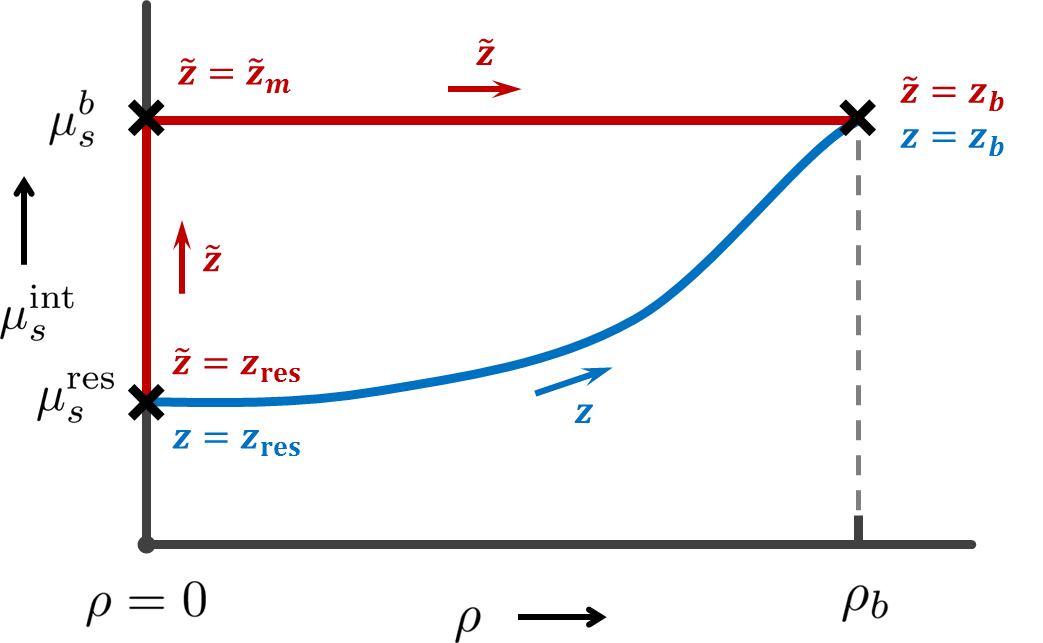}
 \caption{Paths in $(\rho,\mu_s^{\text{int}})$ space.  The physically realized path (blue) is parameterized by $z$ and the path proposed to calculate the integral (\ref{eqn:CrucialIngredient}) (red) by $\tilde{z}$ . The marked points $(\rho=0,\mu_s^{\text{res}})$ and $(\rho_b,\mu_s^b)$ correspond to the reservoir at $z = \tilde{z} = z_{\text{res}}$ and the bulk suspension at $z = \tilde{z} = z_b$  respectively (in the text the bulk density $\rho_b$ is simply called $\rho$). The marked point $(\rho=0,\mu_s^{\text{b}})$ at $\tilde{z} = z_m$ corresponds to a region large enough for a local density approximation to apply.}
 \label{fig:Path}
\end{figure}
\noindent To calculate the remaining integral in (\ref{eqn:TowardsPiI}), we note that the $z$-coordinate parameterizes a path ($\rho(z)$,$\mu_s^{\text{int}}(z)$) in ($\rho$,$\mu_s^{\text{int}}$)-space, as illustrated in blue in Fig. \ref{fig:Path}. This path corresponds to the physically realized profiles $\rho(z)$ and $\mu_s^{\text{int}}(z)$. The crucial observation is that the three pressure terms in Eq. (\ref{eqn:TowardsPiI}) are functions of only $(\rho,\mu_s^b)$ and $(\rho = 0,\mu_s^{\text{res}})$ (at fixed $v_0$), i.e. only of the \emph{endpoints} of the path. Therefore, the same holds true for the integral. This implies that the integral yields the same value when evaluated for a \emph{different} path with the \emph{same} endpoints. The path traced out in $(\rho,\mu_s^{\text{int}})$-space can be altered by applying a nonzero external potential $V_s(z)$ (and thus an external force $\mb{f}^e_s(z) = -\rho_s(z)\partial_zV_s(z)\mb{\hat{z}}$) to the solvent, for according to Eq. (\ref{eqn:IntFBS2}) this alters the profiles $\rho(z)$ and $\mu_s^{\text{int}}(z)$. In particular, we could apply an external potential $V_s(z)$ that vanishes in both the reservoir and the bulk, but is nonzero in between in such a way that the intrinsic solvent chemical potential $\mu_s^{\text{int}}(z)$ increases from $\mu_s^{\text{res}}$ at $z=z_{\text{res}}$ to $\mu_s^b$ already at the reservoir side of the membrane (where $\rho(z)=0$), and such that it remains $\mu_s^{\text{int}}(z) = \mu_s^b$ from there to $z=z_b$ in the bulk suspension on the other side. This situation corresponds to the red path in Fig. \ref{fig:Path}, parameterized by $\tilde{z}$. The marked point $(\rho=0,\mu_s^b)$ along this path corresponds to the region just on the reservoir side of the membrane, where $\rho(\tilde{z})=0$ and $\mu_s^{\text{int}}(\tilde{z}) = \mu_s^b$. We can choose the external solvent potential $V_s(z)$ in such a way that this region is large enough to be considered as a bulk. Upon naming one $\tilde{z}$-coordinate in this intermediate bulk $\tilde{z}_m$ (see Fig. \ref{fig:Path}), the red path can be utilized to calculate the remaining integral in Eq. (\ref{eqn:TowardsPiI}) as
\begin{flalign*}
 \label{eqn:CrucialIngredient}
  \int_{z_{\text{res}}}^{z_b} \mathrm{d}z \rho_s(z) \partial_z \mu_s^{\text{int}}(z)
   = \ \ &\int_{z_{\text{res}}}^{z_b} \mathrm{d}\tilde{z} 
        \rho_s(\tilde{z}) \partial_z \mu_s^{\text{int}}(\tilde{z})
   = \int_{z_{\text{res}}}^{\tilde{z}_m} \mathrm{d}\tilde{z} 
        \rho_s(\tilde{z}) \partial_z \mu_s^{\text{int}}(\tilde{z})\\
   \overset{(\ref{eqn:GDI})}{=} 
    &\int_{z_{\text{res}}}^{\tilde{z}_m} \mathrm{d}\tilde{z} 
     \partial_{\tilde{z}} P_{\text{tot}}(\tilde{z}) \\
   = \ \ &P_{\text{tot}}(\rho=0,\mu_s^b,v_0) 
      - P_{\text{tot}}(\rho=0,\mu_s^{\text{res}},v_0)\\
   \equiv \ \ &\Delta P_s,
\numberthis&&
\end{flalign*}
\noindent where in the second step we used that $\partial_{\tilde{z}} \mu_s^{\text{int}}(\tilde{z}) = 0$ for $\tilde{z}_m \leq \tilde{z} \leq z_b$, and in the third step that $P(\tilde{z})=0$ for $z_{\text{res}} \leq \tilde{z} \leq \tilde{z}_m$, as there are no colloid on the reservoir side of the membrane. As the quantity $\Delta P_s$ defined by Eq. (\ref{eqn:CrucialIngredient}) is a difference in total pressures at colloid density $\rho=0$, it is natural to think of it as a difference in solvent pressures. Indeed, in the limit of a dilute suspension this definition reduces to the difference $\Delta P_s$ used in the main text. To see this, note that the total pressure tensor of Eq. (\ref{eqn:PtotI}) at colloid density $\rho=0$ coincides with the solvent pressure tensor of Eq. (\ref{eqn:SPressureTensor}) (where $\psi^{(2),\text{eq}}_{ss}$ can be replaced by $\psi^{(2)}_{ss}$, as we are considering a vanishing solvent flow, such that $\eta^{(V)}\mbs{\nabla}^2\mb{u}=\mb{0}$).\\
	\indent Combining Eq. (\ref{eqn:PwallI}), (\ref{eqn:TowardsPiI}) and (\ref{eqn:CrucialIngredient}), shows the osmotic pressure to be
\begin{flalign*}
 \label{eqn:PiI}
  \Pi = P(\rho,\mu_s^b,v_0) + \Delta P_s,
\numberthis&&
\end{flalign*}
\noindent which is the extension of the main text's result (\ref{eqn:ActiveHoff}) to a system with interactions.
\noindent Eq. (\ref{eqn:PiI}) shows that the osmotic pressure of a passive system $\Pi = P(\rho,\mu_s^b = \mu_s^{\text{res}},v_0)$ increases with activity in two ways. Firstly, the effective colloids-only pressure $P(\rho,\mu_s^b,v_0)$ of Eq. (\ref{eqn:PressureTensorOverdamped}) changes, as the two-body colloid-colloid correlation function changes with activity \cite{Speck2016EffectivePairPotential}, and as the increasing solvent chemical potential may modify the effective colloid-colloid interaction potential and the colloid-colloid correlation function. Secondly, the increase in solvent chemical potential induces an increase $\Delta P_s$, that can be interpreted as an increase in solvent pressure.
\section{4. Derivation of the solvent pressure difference $\Delta P_s$ (with interactions)}
\label{sec:DeltaPs}
\noindent The solvent pressure difference $\Delta P_s$ can be expressed as
\begin{flalign*}
 \label{eqn:DeltaPs2}
   \Delta P_s 
  \overset{(\ref{eqn:CrucialIngredient})}{=}
   \int_{z_{\text{res}}}^{z_b}\mathrm{d}z
   \rho_s(z)\partial_z\mu_s^{\text{int}}(z)
  \overset{(\ref{eqn:IntFBS2})}{=}
   -\int_{z_{\text{res}}}^{z_b}\mathrm{d}z
   f^p_z(z)
  =
   -\gamma_tv_0
   \int_{z_{\text{res}}}^{z_b} \mathrm{d}z 
   m_z(z),
\numberthis&&
\end{flalign*}
\noindent where we used $\mb{f}^p(z) = f^p_z(z) \hat{\mb{z}} = \gamma_t v_0 m_z(z)\mb{\hat{z}}$, and where we assumed the membrane to exert no force on the solvent, i.e. $\mb{f}_s^e(\mb{r})=0$. In the dilute limit, the same expression follows alternatively from finding $\Delta P_s = P_s(z_b) - P_s(z_{\text{res}})$ from Eq. (\ref{eqn:FBS}) of the main text (with $\mb{f}^e_s(\mb{r})=\mb{f}^f(\mb{r})=\eta\mbs{\nabla}^2\mb{u}(\mb{r})=0$).
\noindent An expression for the polarization $m_z(z)$ is found from the Smoluchowski equation, formed by inserting Eq. (\ref{eqn:Currents}) into Eq. (\ref{eqn:OneBodyM0}). Indeed, the first moment (in the variable $\mb{\hat{e}}$) of this Smoluchowski equation yields, in steady state, and in the same geometry as in the main text,
\begin{flalign*}
 \label{eqn:PolEqn}
 \begin{alignedat}{2}
   2\frac{k_BT}{\gamma_r} m_z(z) 
  = 
   &-\partial_z
         \bigg\{ v_0\left(\frac{1}{3}\rho(z) + \St_{zz}(z)\right)
                &&-\gamma_t^{-1} \int\mathrm{d}\mb{\hat{e}}\psi(z,\theta) e_z \partial_z V(z,\theta)
                -\gamma_t^{-1}k_BT \partial_z m_z(z) \\
                &&&-\gamma_t^{-1}
                  \int\mathrm{d}\mb{\hat{e}}\mathrm{d}\mb{r'}\mathrm{d}\mb{\hat{e}'}
                  e_z\partial_z\phi^{\text{eff}}_{\mb{\hat{e}}\mb{\hat{e}'}}(\mb{r'}-\mb{r})
                  \psi^{(2)}_{\mb{\hat{e}}\mb{\hat{e}'}}(\mb{r},\mb{r'})
         \bigg\}\\
      &\ \mathrlap{
        +\gamma_r^{-1}
          \int\mathrm{d}\mb{\hat{e}} 
          \sin(\theta) \partial_{\theta}V(z,\theta)
          \psi(z,\theta)
        +\gamma_r^{-1}
          \int\mathrm{d}\hat{\mb{e}}\mathrm{d}\mb{r'}\mathrm{d}\mb{\hat{e}'}
          \sin(\theta)\partial_{\theta}\phi^{\text{eff}}_{\mb{\hat{e}}\mb{\hat{e}'}}(\mb{r'}-\mb{r})
          \psi^{(2)}_{\mb{\hat{e}}\mb{\hat{e}'}}(\mb{r},\mb{r'}),
       }
 \end{alignedat}
 \numberthis&&
\end{flalign*}
\noindent where $e_z = \cos\theta$ and where $\St(z) \equiv \int\mathrm{d}\mb{\hat{e}} \psi(z,\theta)\left(\mb{\hat{e}}\mb{\hat{e}}-\frac{1}{3}\Id\right)$ is the traceless alignment tensor in three dimensions. According to Eq. (\ref{eqn:DeltaPs2}), the difference in solvent pressure $\Delta P_s$ essentially follows from integrating Eq. (\ref{eqn:PolEqn}) from the reservoir to the bulk suspension. The first two lines of the right-hand side of Eq. (\ref{eqn:PolEqn}) are easily integrated, as they form a derivative with respect to $z$. Note that all the terms acted upon by the derivative vanish in the reservoir, whereas only the density term and the interaction term are nonzero in the bulk suspension. The resulting solvent pressure difference is
\begin{flalign*}
 \label{eqn:DeltaPs3}
   \Delta P_s 
  = 
  &\frac{\gamma_t\gamma_rv_0^2}{6k_BT}\rho
   -\frac{\gamma_rv_0}{2k_BT}
     \int\mathrm{d}\mb{\hat{e}}\mathrm{d}\mb{r'}\mathrm{d}\mb{\hat{e}'}
     e_z\partial_z\phi^{\text{eff}}_{\mb{\hat{e}}\mb{\hat{e}'}}(\mb{r'}-\mb{r}_b)
     \psi^{(2)}_{\mb{\hat{e}}\mb{\hat{e}'}}(\mb{r}_b,\mb{r'})\\
  &-\frac{\gamma_tv_0}{2k_BT}
     \int_{z_{\text{res}}}^{z_b}\mathrm{d}z
     \int\mathrm{d}\mb{\hat{e}}
     \sin(\theta)\partial_{\theta}V(z,\theta)
     \psi(z,\theta)\\
  &-\frac{\gamma_tv_0}{2k_BT}
     \int_{z_{\text{res}}}^{z_b}\mathrm{d}z
     \int\mathrm{d}\mb{\hat{e}}\mathrm{d}\mb{r'}\mathrm{d}\mb{\hat{e}'}
     \sin(\theta)\partial_{\theta}\phi^{\text{eff}}_{\mb{\hat{e}}\mb{\hat{e}'}}(\mb{r'}-\mb{r})
     \psi^{(2)}_{\mb{\hat{e}}\mb{\hat{e}'}}(\mb{r},\mb{r'}),
 \numberthis &&
\end{flalign*}
\noindent where $\mb{r}_b$ is a point in the bulk suspension. In the dilute limit (where the interaction terms are negligible), this solvent pressure difference reduces to Eq. (\ref{eqn:DeltaPs}), as was claimed in the main text.
\section{5. Numerical solution details}
\label{sec:Numerics}
\begin{figure}[t]
 \includegraphics[width=0.5\linewidth]{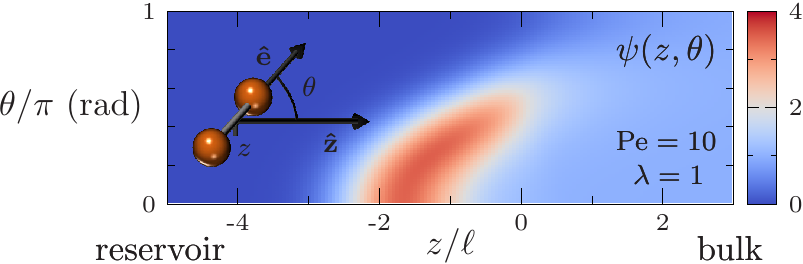}
 \caption{Normalized probability $\psi(z,\theta)$ to find a dumbell at (scaled) position $z$ with orientation $\theta$, obtained as a numerical solution to Eq. (\ref{eqn:FP}) of the main text for a soft potential $V(z)$ at $z<0$ acting on both halves of the dumbell (see text), with potential strength $\lambda$ and activity Pe as indicated. Note that $\psi(z<-4,\theta) \approx 0$, indicating that (almost) no particles penetrate into the wall beyond $z=-4$, and that $\psi(z>3,\theta) \approx 1$, indicating a homogeneous and isotropic bulk for $z>3$. The probability peak corresponds to dumbells persistently propelling into the membrane.}
 \label{fig:Psi}
\end{figure}
\noindent The plots in figures \ref{fig:IsoHarm} and \ref{fig:Osmo2} are based upon a numerical solution $\psi(z,\theta)$ to the Smoluchowski equation (\ref{eqn:FP}), in steady state in the planar geometry of the main text. From this solution the other quantities of interest (e.g. polarization and pressure profiles) follow. The solutions $\psi(z,\theta)$ were obtained, using COMSOL Multiphysics$^{\textregistered}$, for $z$-values $-20 \ell \leq z \leq 20 \ell$, where $\ell \equiv \sqrt{\gamma_r/\gamma_t}$, and $\theta$-values $0 \leq \theta \leq \pi$. As Fig. \ref{fig:IsoHarm} corresponds to spherical particles, the considered colloid-membrane interaction potential, $V(z) = \lambda k_BT (z/\ell)^2$ with $\lambda = 1$ for $z<0$ and $V(z)=0$ for $z \geq 0$, is independent of $\theta$. Fig. \ref{fig:Osmo2} corresponds to dumbells, consisting of two point particles with separation $\ell$, both of which are subject to the same potential $V(z)$, where $\lambda$ is varied.
The potential experienced by one dumbell, with center of mass position $z$ and orientation characterized by the polar angle $\theta$, follows as $V(z,\theta) = V\left(z - \ell \cos\theta/2\right) + V\left(z + \ell\cos\theta/2\right)$.
The applied boundary conditions are in both cases (i) a uniform distribution in the bulk, i.e. $\psi(20\ell,\theta) = c$ with arbitrary normalization $c$, (ii) either no flux $j_z(z=-20\ell,\theta) = 0$ or no particles $\psi(z=-20\ell,\theta) =0$ far into the quadratic potential (both conditions yield equivalent solutions), and (iii) $\partial_{\theta}\psi(z,0) = 0$ and $\partial_{\theta}\psi(z,\pi)= 0$, which follow from the symmetries $\psi(z,\theta) = \psi(z,-\theta)$ and $\psi(z,\pi + \theta) = \psi(z,\pi-\theta)$, respectively. A typical solution $\psi(z,\theta)$ for dumbells, shown in Fig. \ref{fig:Psi}, displays the same physics as encountered for spheres in the main text, namely an accumulation of dumbells at the membrane caused by dumbells persistently propelling into the membrane.
\section{6. Solvent flow in a pipe}
\noindent This section illustrates that the net polarization of colloids near a semipermeable membrane can also lead to solvent flow. Consider a dilute, active suspension in a cylindrical pipe, of radius $R$ and length $L$, confined on one side by a semipermeable membrane, as illustrated in Fig. \ref{fig:PipeFlow} of the main text. The difference with the setup of Fig. \ref{fig:NicePicture} of the main text is in the boundary conditions: whereas Fig. \ref{fig:NicePicture} corresponds to no-flux boundary conditions, the boundary conditions instead imposed here are equal solvent pressures on either end. We adopt a cylindrical coordinate system, where the $z$-axis coincides with the symmetry axis of the cylinder, $z=\pm L/2$ corresponding to either end of cylinder, where $r$ is the radial distance from the $z$-axis, $\mb{\hat{r}}$ being the corresponding unit vector, and where $\phi$ is the azimuthal angle. We shall assume the solvent velocity $|\mb{u}(\mb{r})| \ll v_0$, such that the effect of advection is negligible. The dynamics of the colloids is then governed by the Smoluchowski equation (\ref{eqn:FP}) of the main text. We again consider a steady state where the colloid flux vanishes (such that $\mb{f}^f(\mb{r})=\mb{0}$). We then expect the colloids to accumulate at, and form a net polarization towards, both the membrane and the outer walls of the pipe. We assume the radius $R$ to be large, i.e. $R \gg \sqrt{\gamma_r/\gamma_t}$ and $R \gg \gamma_r v_0/k_BT$, such that in the centre of the pipe the effect of the pipe's outer wall is not felt. In this centre region, that we denote as $r < R^* < R$, the results for the planar geometry of the main text then carry over. More precisely, we expect the propulsion force $\mb{f}^p(r,z) = f^p_z(r,z) \mb{\hat{z}} + f^p_r(r,z)\mb{\hat{r}}$ to be given by $\mb{f}^p(r,z) = f^p_z(z)\mb{\hat{z}}$ for $r < R^*<R$, where $f_z^p(z)$ describes the polarization profile next to a membrane in a planar geometry. In particular, $-\int_{-L/2}^{L/2}\mathrm{d}zf^p_z(z) = \Delta P_s$, as calculated in Eq. (\ref{eqn:DeltaPs}) of the main text. Note that the interpretation of $\Delta P_s$ as a solvent pressure difference is $\emph{not}$ valid in this setting, for this interpretation relies on the solvent flow $\mb{u}(\mb{r}) = 0$. Instead, at this point $\Delta P_s$ should purely be regarded as the right-hand side of Eq. (\ref{eqn:DeltaPs}). Finally, we assume the colloid polarization to be such that $\mbs{\nabla} \times \mb{f}^p(r,z) = \mb{0}$, also for $R^* \leq r \leq R$. The steady state solvent velocity profile $\mb{u}(r,z)$ is then governed by the Stokes equation (\ref{eqn:FBS})
\begin{flalign*}
 \label{eqn:StokesFlow}
  - \mb{f}^p - \mbs{\nabla}P_s + \eta \mb{\nabla}^2 \mb{u}  = 0,
 \numberthis&&
\end{flalign*}
\noindent which again features the opposite propulsion force $-\mb{f}^p$ as a body force (note that we used $\mb{f}^e_s=\mb{f}^f=\mb{0}$). Eq. (\ref{eqn:StokesFlow}) is to be solved, together with the incompressibility condition $\mbs{\nabla} \cdot \mb{u} = 0$, subject to the boundary conditions of equal solvent pressure $P_s(r,z=-L/2) = P_s(r,z=L/2) = P_0$ for $r < R^*$, and the no-slip condition $\mb{u}(r=R,z) = \mb{0}$.
Upon using the ansatz $\mb{\hat{r}} \cdot \mb{u} = 0$, the incompressibility condition reveals that $\mb{u}(r,z) = u_z(r)\mb{\hat{z}}$. Inserting this into (\ref{eqn:StokesFlow}) yields
\begin{subequations}
 \label{eqn:TowardsPFlow}
 \begin{flalign*}
  &- f^p_r -\partial_r P_s = 0, 
    \numberthis\label{eqn:TowardsPFlowa}\\
  &- f^p_z - \partial_zP_s + \eta \frac{1}{r} \partial_r (r\partial_r u_z) = 0.
  \label{eqn:TowardsPFlowb}
 \numberthis &&
 \end{flalign*}
\end{subequations}
\noindent Deriving Eq. (\ref{eqn:TowardsPFlowb}) with respect to $r$, and using $\mbs{\nabla}\times \mb{f}^p=\mb{0}$ together with Eq. (\ref{eqn:TowardsPFlowa}), shows that $\eta \frac{1}{r}\partial_r(r\partial_ru_z(r))=c$. The constant $c$ follows by integrating (\ref{eqn:TowardsPFlowb}) from $z=-L/2$ to $z=L/2$ (for $r<R^*$), resulting in $c=-\Delta P_s/L$. Solving for $u_z(r)$, and using the no-slip boundary condition, yields the velocity profile
\begin{flalign*}
 \label{eqn:PFlow}
  u_z(r) = \frac{\Delta P_s}{4 \eta L} (R^2 - r^2) \mb{\hat{z}}.
 \numberthis &&
\end{flalign*}
\noindent Eq. (\ref{eqn:PFlow}) shows that the solvent flow - driven by the body force $-\mb{f}_p$ - coincides with a Poiseuille-Hagen flow driven by a solvent pressure difference $\Delta P_s$ applied between the ends of the pipe.

\def\bibsection{\section{7. ESI References}} 
\putbib[../../../../Literature/Database/ActiveMatter/ActiveMatter]
\end{bibunit}

\end{document}